%% file: main.tex
\documentclass[twocolumn]{article}

\usepackage{bttda-paper}
\usepackage{todonotes}

\title{BTTDA: Block-Term Tensor Discriminant Analysis for Brain-Computer Interfacing}

\author{%
	A. Van Den Kerchove$^{1,2,*}$,
	H. Si-Mohammed$^{2}$,
	F. Cabestaing$^{2}$,
	M.M. Van Hulle$^{1}$
	\bigskip\\
	$^1$ KU Leuven,
	Leuven Brain Institute,
	Leuven.AI,\\
	Laboratory for Neuro- and Psychophysiology,	\\
	Campus Gasthuisberg O\&N2,
	Herestraat 49 bus 1021,
	BE-3000 Leuven,
	Belgium
	\smallskip\\
	$^2$ Univ. Lille, CNRS, Centrale Lille,
	UMR 9189 CRIStAL,
	F-59000 Lille,
	France
	\smallskip\\
	$^*$ \texttt{arne.vandenkerchove@kuleuven.be}
}

\begin{document}

\maketitle

\begin{abstract}
	\input{abstract.txt}

\end{abstract}

\section{Introduction}

\Acp{bci} have the potential to bypass
defective neural pathways by providing an alternative communication channel
between the brain and an external device.
These interfaces find applications in the development of
neuroprosthetics, assistive technologies and rehabilitation~\cite{Wolpaw2020}.
To achieve their functionality, \acp{bci} record and process neural data,  with
\ac{eeg} the most popular recording method in the field.

A \ac{bci} usually operates by identifying specific, task-related activity in
the recorded \ac{eeg} data, which can then be coupled to output or actions.
This methodology often gives rise to classification problems~\cite{Lotte2018}.
Some well-known examples include the P300 speller~\cite{Krusienski2006}, where
momentary visual stimuli evoke characteristic \acp{erp} modulated by attention,
and \ac{mi}~\cite{Aggarwal2019}, where different (imagined) limb movements evoke
\acp{ersd} with different spatial patterns.
As a consequence \ac{bci} decoding (\ac{erp} vs.\ non-attended \ac{erp}, left
vs.\ right limb \ac{mi}, \ldots) involves a calibration phase training a classifier
on labeled \ac{eeg} data and an operation phase where the trained classifier is applied
to unseen \ac{eeg} data.

The duration of the calibration session should ideally be minimized to enhance
user experience.
This results in small, subject- and session-specific training datasets
which make \ac{bci} classification methods vulnerable to overfitting in the
presence of high-dimensional data.
One possible countermeasure is applying a dimensionality reduction technique
which extracts a lower-dimensional set features relevant to the classification
problem at hand.

\subsection{Tensors \& tensor methods}

Because of the multichannel time series format of \ac{eeg} and other \ac{bci} functional
neuroimaging methods, recorded data naturally exist as multiway data, capturing
information in both the spatial and the temporal domain.
Preprocessing transformations can further expand the data into additional
analytic domains.
Common examples include time-frequency transformation, time-binning, or
integrating information across multiple subjects or conditions.
This in turn results in high-dimensional datasets which are usually flattened
into a set of sample vectors, stripping the original data of its structure.
A more suited approach relies on this intrinsic multiway structure of neural
data~\cite{Erol2022} to represent the data as \emph{tensors}, multiway arrays,
with each domain corresponding to a tensor \emph{mode}.
Tensors provide a structured data representation for this highly dimensional
multiway data.
This in turn paves the way to the development of tensor methods which can
counteract some of the drawbacks of the dimensionality problem.
Tensor methods are machine learning or dimensionality reduction techniques that
consider each tensor mode separately, reducing a given problem into partial,
per-mode problems.

\begin{figure*}[t]
	\centering
	\makebox[\linewidth][c]{%
		\input{tensor_core_structures.tikz.tex}
	}
	\caption[Core tensor(s) of different tensor decomposition structures.]{%
		A tensor decomposition finds core tensor and factor matrices
		from input tensor.
		This core tensor can have several structures.
		In the Tucker structure, the core is a dense tensor $\ten{G}$.
		The \ac{parafac} structure expresses the core as a sum of $B$ rank-1 terms, each
		with a scalar core $g^{(b)}$.
		The block-term structure expresses the core as a sum of $B$ smaller,
		Tucker-structured blocks $\ten{G}^{(b)}$.
		Both the \ac{parafac} and block-term structures are more sparse than the full
		Tucker structure, yet the block-term structure is more flexible as it
		allows blocks of variable tensor dimensionality instead of fixed rank-1 terms.
	}\label{fig:bttda/sparse}%
\end{figure*}
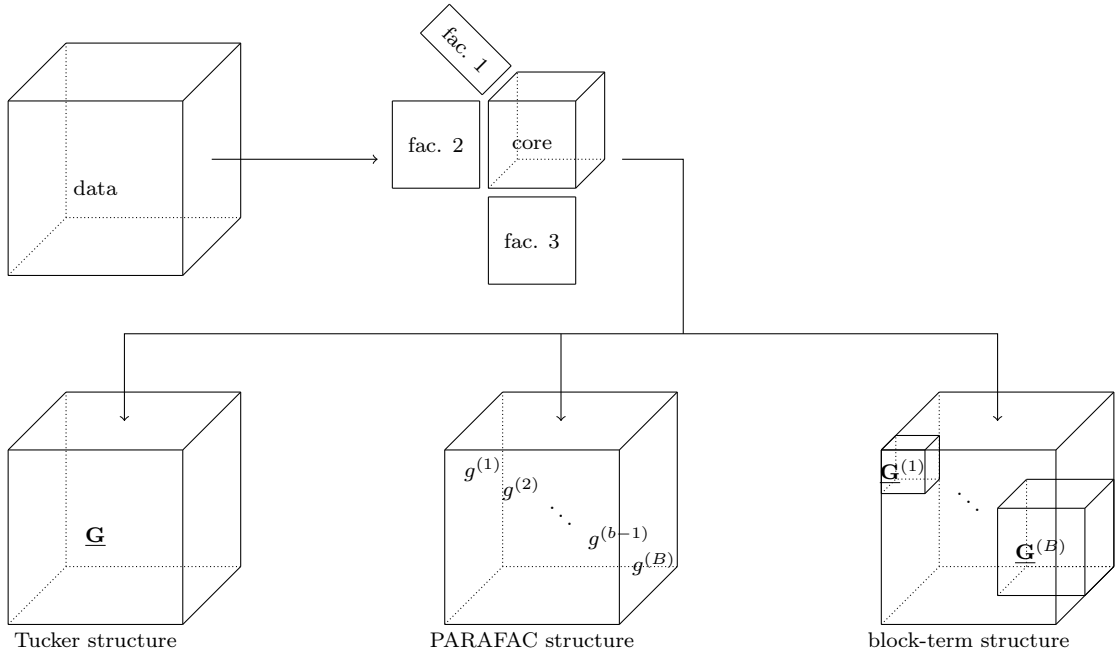

Tensor methods often decompose tensors into a lower dimensional structure
of a core tensor and factor tensors.
The most common approaches adhere to either the Tucker structure or the PARAFAC
structure.
A Tucker decomposition reduces an input tensor of order $K$ with dimensions
$(D_1,D_2,\ldots,D_K)$ to a dense tensor with dimensions $(R_1,R_2,\ldots,R_K)$,
with $R_k \leq D_k$ for $k=1, 2, \ldots K$, using a set of per-mode factor
matrices.
Effective unsupervised tensor decomposition and approximation in the Tucker format can be achieved
using the \ac{hosvd}~\cite{DeLathauwer2000,SoleCasals2018}.
Alternatively, the \ac{parafac} structure can be used.
Here, the tensor is decomposed into a sum of rank-1 tensors, each the product
of a scalar and a vector per mode.
This is equivalent to a Tucker structured decomposition with all core elements
off the hyperdiagonal set to 0, as shown in \cref{fig:bttda/sparse}.
One way of obtaining an unsupervised PARAFAC decomposition is through the Canonical Polyadic
Decomposition~\cite{Hitchcock1927,Nazarpour2006}.
These decomposition methods can be regarded as feature extraction methods for a
\ac{bci} classification problem, with the flattened core tensors as feature vectors.
Extracted features can subsequently be classified to predict class labels, most
commonly using \ac{lda} or a \ac{svm}.

While commonly used, these Tucker or \ac{parafac} structures  might still not be able to
efficiently represent relevant neural information in a compressed format.
The block-term tensor structure is a generalization of both the Tucker and
\ac{parafac} structures.
It represents the tensor as a sum of Tucker structured terms.
If the number of terms is equal to 1, it is equivalent to the Tucker structure;
if the dimensions of each term are equal to 1, it is equivalent to the \ac{parafac}
structure.
The block-term structure (\cref{fig:bttda/sparse}, right) is more flexible than
either the Tucker or the \ac{parafac} structures, since it is not constrained to
solutions that must be expressed as either one of these structures and their
chosen hyperparameters.
Due to its flexibility, the block-term structure can strike a better
balance between extracting a maximal amount of relevant features and a minimal
amount of irrelevant features.
However, this increased flexibility comes at the cost of a higher number of
hyperparameters, as now both the number of terms and the dimension of
each term need to be specified.
A block-term structured core tensor can be obtained in an unsupervised way using
the \ac{btd}~\cite{DeLathauwer2008,DeLathauwer2008a,DeLathauwer2008b,Rontogiannis2021}.
Performance of methods leveraging either the Tucker and \ac{parafac} structures are
heavily dependent on the prior choice of hyperparameters describing
the desired reduced dimension or the number of rank-1 terms.

\subsection{Supervised tensor decompositions for \ac{bci}}

If the decompositions are not full rank, the Tucker, \ac{parafac} and block-term
structures are not unique and can be obtained by optimizing different criteria.
Given the low signal-to-noise ratio and specific, task-related output expected
in a \ac{bci} application, supervised feature extraction and machine learning techniques are
favored~\cite{Lotte2018} over the unsupervised decomposition methods presented
above.
A decomposition that is helpful for classification should ideally optimize
the discriminability between classes in the resulting core tensors, which can
then be considered as extracted features.
In this philosophy, the Tucker decomposition can also be obtained
using \ac{hoda}~\cite{Yan2005,Phan2010,Froelich2018}, which optimizes class
separability in the Fisher sense, analogous to linear discriminant analysis.

Variants of \ac{hoda} have been applied to \ac{bci} problems such as the decoding
of \acp{erp}~\cite{Onishi2012,Higashi2016} and \ac{mi}~\cite{Liu2015,Cai2021}
with positive results~\cite{Lotte2018}.
Recent work proposes optimization of the objective
function and introduces regularization~\cite{JamshidiIdaji2017,Jorajuria2022,Aghili2023}.
Discriminant tensor features have also been extracted
in the \ac{parafac} structure through manifold optimization~\cite{Froelich2018}.
However, it is not immediately obvious if either the Tucker or \ac{parafac}
structure are most suited to represent the neural data of interest for the
\ac{bci}
paradigm and for decoding.

More recent studies have shown that supervised decoders adopting a more flexible structure
can improve \ac{bci} performance.
Promising results have been achieved for regression tasks using
\ac{hopls}~\cite{Zhao2012,Camarrone2018} and \ac{bttr}~\cite{Faes2022,Faes2022a}.
\ac{bttr} has also been adapted into the classification variant \ac{bttc}~\cite{Camarrone2021}
but this methodology leaves room for improvement:
instead of optimizing features directly for class separability, \ac{bttc} regresses
to dummy 2-valued independent variable.
Thus, the method cannot be extended to a multi-class setting.
Furthermore, structures employed in these regression approaches could still
be considered as relatively constrained in comparison.
A more flexible approach could rely on a full block-term tensor decomposition
of the input data which optimizes discriminability and relies on a low-rank
common subspace between the input and classification labels.
\textcite{Huang2020} propose a supervised approach for finding multiple discriminant
multilinear spectral filter terms and apply it to motor imagery BCI, but their
decomposition is also limited in flexibility, since the solution is
restricted to terms with  dimension $(R_1,R_2,1)$, with mode 3
corresponding to the frequency domain.

\subsection{Contribution: A block-term structured model for classification}

With a proper choice of reduced dimension and number of terms, a
block-term decomposition directly optimizing discriminability might be more
suited to represent complex neural data in a sparse way, which additionally
yields a regularization effect.
Multiple parsimonious discriminant block terms with lower
dimensions might yield better performance than a single \ac{hoda} block
requiring a higher dimension to capture discriminant information, and by doing
so extracts too many irrelevant features.
A complementary view on the same approach goes as follows:
if HODA with a well-chosen reduced dimension extracts some discriminant features
from the input tensor, it is likely that it does not retrieve all useful
information due to the restrictions imposed by its Tucker structure.
Could \ac{hoda} therefore not sequentially be applied to extract discriminant
Tucker structured terms -- potentially with lower dimension -- as long as decoding
performance increases?

We implement this idea as a novel supervised feature
extraction method titled \ac{bttda}, a generalization of the aforementioned
\ac{hoda} algorithm.
\Ac{bttda} extracts discriminant features while adhering to a
flexible and efficient block-term tensor structure.
This work features the following contributions:
\begin{enumerate*}[label={\arabic*)}]
	\item We develop a forward model for \ac{hoda} to reconstruct a
	      given input tensor from the extracted features.
	\item This allows us to introduce \ac{bttda} as a state-of-the-art \ac{bci}
	      feature extraction method based on the block-term tensor structure.
	\item We evaluate a \ac{bci} decoder based on \ac{bttda} and its special
	      \ac{parafac}-structured case on decoding benchmarks for both \ac{erp}
	      and \ac{mi}
	      \ac{bci} paradigms and compare these to state-of-the-art decoders.
\end{enumerate*}

\section{Methods}

\subsection{Notation}
Tensors are indicated by bold underlined letters $\ten{X}$, matrices by bold
letters $\mat{U}$, fixed scalars by uppercase letters $K$, and variable
scalars as lowercase letters $k$.
The $n^\text{th}$ sample of a tensor dataset with $N$ samples is written as
$\ten{X}(n)$, the dataset itself as ${\{\ten{X}(n)\}}_n^N$.
A tensor $\smash{\ten{X}\in \mathbb{R}^{D_1\times D_2 \times \cdots \times D_K}}$ can be
unfolded in mode $k$ to a matrix
$\smash{\mat{X}_k\in\mathbb{R}^{(D_k\times\prod_{j\neq k}^K D_j)}}$, by concatenating
all mode $j\neq k$ fibers.
The tensor-matrix product of tensor $\ten{X}$ with matrix $\mat{U}$ along a
given mode $k$ is written as $\ten{X}\mpr{\mat{U}}{k}$. For ease of notation, let
$\ten{X}\mmpr{\mat{U}} =
	\ten{X}\mpr{\mat{U}}{1}\mpr{\mat{U}}{2}\cdots\mpr{\mat{U}}{K}$.
When skipping one of the modes $k$, this is
written as $\ten{X}\mmprs{\mat{U}}{k} =
	\ten{X}\mpr{\mat{U}}{1}\mpr{\mat{U}}{2}\cdots\mpr{\mat{U}}{k-1}\mpr{\mat{U}}{k+1}\ldots\mpr{\mat{U}}{K}$.
$\mat{A}\otimes\mat{B}$ indicates the Kronecker product of matrices $\mat{A}$
and $\mat{B}$.

\subsection{\Acl{hoda}}
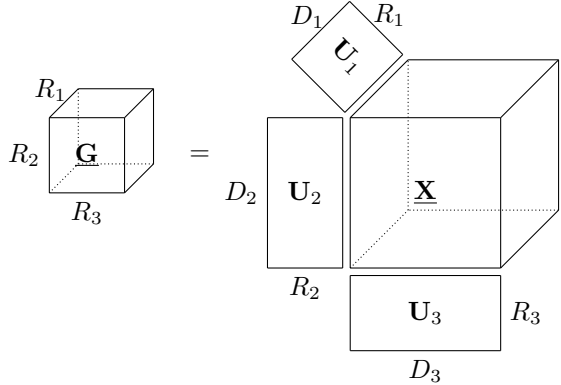
\begin{figure}[t]
	\centering
	\input{hoda_bw.tikz.tex}
	\caption[A \acs{hoda} backward projection.]{%
		A visualization of the multilinear projection obtained by \acf{hoda} applied to a third-order tensor
		sample $\ten{X}$ with dimensions $(D_1,D_2, D_3)$.
		\Ac{hoda} finds projection matrices $\mat{U}_k$ such that maximal
		discriminability between classes is achieved in the projected latent tensors
		$\ten{G}$ with reduced dimension $(R_1,R_2,R_3)$.}
	\label{fig:hoda-backward}%
\end{figure}

\Acf{hoda}~\cite{Phan2010} is a
supervised, tensor-based dimensionality reduction and feature extraction technique.
For a set of $N$ tensors of order $K$
$\left\{\ten{X}(n)\in\mathbb{R}^{D_1\times D_2 \times \cdots \times
		D_K}\right\}_n^N$, HODA finds projection matrices $\mat{U_k}$ for each mode $k$
which project a given $\ten{X}$ to a latent tensor
$\ten{G}\in\mathbb{R}^{R_1\times R_2\times\cdots\times R_K}$, usually with lower
dimensions $(R_1\leq D_1,R_2\leq D_2,\ldots,R_K\leq D_K)$ using
tensor-matrix mode products:
\begin{equation}
	\ten{G}  = \ten{X}\mmpr{\mat{U}}
	\label{eq:HODA-backward}
\end{equation}
as visualized in \cref{fig:hoda-backward}.
Since \ac{hoda} extracts latent features or properties $\ten{G}$ from the observed data
$\ten{X}$, relying on a task-related criterion, it can be referred to as a
\emph{backward model}.

Analogous to the \ac{hosvd}, \ac{hoda} decomposition results in a dense latent
tensor $\ten{G}$ and imposes an orthogonality constraint on each $\mat{U}_k$ to ensure uniqueness.
However, while \ac{hosvd} projection matrices minimize the reconstruction error,
\ac{hoda} optimizes the class discriminability of the reduced tensors
$\ten{G}(n)$ belonging to classes with labels $c_n$.
This is a desirable property in a classification setting where samples are
high-dimensional tensors.

\Ac{hoda} optimizes discriminability in the Fisher sense, maximizing the Fisher
ratio $\phi$ between the latent tensors $\ten{G}(n)$:
\begin{equation}
	\phi\left(\left\{\mat{U}\right\}\right) = \frac{\sum_c^CN_c\left\|\bar{\ten{G}}(c)-\bar{\bar{\ten{G}}}\right\|_F^2}
	{\sum_n^N\left\|\ten{G}(n)-\bar{\ten{G}}(c_n)\right\|_F^2}
	\label{eq:fisher}
\end{equation}
for $C$ classes with each $N_c$ samples. $\bar{\ten{G}}(c)$ is the mean of
latent tensors of class $c$, and $\bar{\bar{\ten{G}}}$ the mean of
these class mean latent tensors.
If the dimensions $(R_1,R_2, \ldots,R_k)$ are set a priori, the objective is now
to find the optimal projection matrices:
\begin{equation}
	\left\{\mat{U}^*\right\} =  \argmax_{\{\mat{U}\}}\phi\left(\left\{\mat{U}\right\}\right)
\end{equation}
which is solved through the backward HODA algorithm.
To start, $\mat{U}_k$ are initialized to orthogonal matrices, e.g.,\ as random
orthonormal matrices, by a per-mode \ac{svd},
or as the partial \ac{hosvd} of all stacked tensors in the dataset.
At each iteration, the algorithm loops through the modes and fixes all
projections but $\mat{U}_k$ corresponding to mode $k$.
It then finds a partial latent tensor:
\begin{equation}
	\ten{G}_{-k}=\ten{X}\mmprs{\mat{U}}{k}
\end{equation}
Subsequently, a new projection matrix $\mat{V}_k$ can be found analogous to Linear
Discriminant Analysis by constructing the partial within-class scatter matrix:
\begin{equation}
	\mat{S}_{-k,\text{w}} = \sum_n^N\tilde{\mat{G}}_{-k,k}(n)\cdot\tilde{\mat{G}}_{-k,k}^\intercal(n)
\end{equation}
with $\tilde{\ten{G}}_{-k}(n) = \ten{G}_{-k}(n) - \bar{\ten{G}}_{-k}(c_n)$,
and the partial between-class scatter matrix:
\begin{equation}
	\mat{S}_{-k,\text{b}} =
	\sum_c^CN_c\tilde{\bar{\mat{G}}}_{-k,k}(c)\cdot\tilde{\bar{\mat{G}}}_{-k,k}^\intercal(c)
\end{equation}
with $\tilde{\bar{\ten{G}}}_{-k}(c) = \bar{\ten{G}}_{-k}(c) - \bar{\bar{\ten{G}}}_{-k}$,
and solving for the $R_k$ leading eigenvectors in the eigenvalue problem:
\begin{equation}
	\mat{S}_{-k,\text{b}}-\varphi_k\mat{S}_{-k,\text{w}} =
	\mat{V}_k\mat{\Lambda}\mat{V}_k^\intercal
\end{equation}
with $\varphi_k=\tr\left(\mat{U}_k^\intercal\mat{S}_{-k,\text{b}}\mat{U}_k\right)/\tr\left(\mat{U}_k^\intercal\mat{S}_{-k,\text{w}}\mat{U}_k\right)$
using the $\mat{U}_k$ obtained in the previous iteration.
Finally, the orthogonal transformation invariant projections $\mat{U}_k$
are obtained by calculating the
per-mode total scatter matrices:
\begin{equation}
	\mat{S}_{k,\text{t}} = \sum_n^N\mat{X}_k(n)\cdot\mat{X}_k^\intercal(n)
\end{equation}
and finding the $R_k$ leading eigenvectors of:
\begin{equation}
	\mat{V}_k\mat{V}_k^\intercal\mat{S}_{k,\text{t}}\mat{V}_k\mat{V}_k^\intercal
	= \mat{U}_k\mat{\Lambda}\mat{U}_k^\intercal
\end{equation}
at each iteration~\cite{Wang2007}.
The iterative process halts when the
update of each $\mat{U}_k$ is lower than a predetermined threshold $\epsilon$ or after a
fixed number of iterations $I_\text{max}$.
The full \ac{hoda} procedure is summarized in \cref{alg:HODA}.
\begin{algorithm}
	\caption[A \acs{hoda} backward solution.]{The \acs{hoda} backward solution.}
	\label{alg:HODA}
	\input{alg_hoda_bw.tex}
\end{algorithm}

To apply \ac{hoda} in a classification setting, the projections
are first learned on a training dataset with known class labels.
Next, these projections are used to extract latent tensors from the
tensors in the training dataset.
These latent training tensors are then reshaped (\emph{vectorized}) into feature vectors
$\mat{g} =  \vect(\ten{G})$ and used to train a decision classifier with the corresponding class labels.
At the evaluation stage, the projections learned from the training dataset are
used to extract latent tensors from an unseen test dataset with unknown class
labels, which can also be vectorized and passed on to the trained decision
classifier.

To avoid overfitting and improve performance in low sample size settings, the
HODA problem can be regularized by shrinking the partial
within-class scatter matrices~\cite{Phan2010} with a shrinkage factor
$\alpha_k$ at each step such that the eigenvalue problem becomes
\begin{equation}
	\mat{S}_b^{(-k)} -
	\varphi\left[\left(1-\alpha_k\right)\mat{S}_{-k,\text{w}}+\alpha_k\mat{I}\right] =
	\mat{V}_k\mat{\Lambda}\mat{V}_k^\intercal
\end{equation}
As in Linear Discriminant Analysis, the shrinkage parameter $\alpha_k$ can
also be estimated in a data-driven way in HODA~\cite{Jorajuria2022},
e.g., using the Ledoit-Wolf procedure~\cite{Ledoit2003} at every iteration.

\subsection{A forward model for \acs{hoda}}

As a prerequisite to the proposed \ac{bttda} model, we must find a
way to reconstruct the original data tensor $\ten{X}$ as accurately as possible
from $\ten{G}$ after dimensionality reduction.
This requires a \emph{forward} model, a generative model that expresses the observed data in
terms of given latent properties or features.
As indicated earlier, finding the optimal projection matrices ${U}$ that extract
tensors $\ten{G}$ given input data $\ten{X}$ as in \cref{eq:HODA-backward}
corresponds to fitting a backward \ac{hoda} model.
A forward model is a method to reconstruct the original data $\ten{X}$
from the core tensor $\ten{G}$.
Forward models are useful for, e.g., interpretability and data compression,
but here reconstruction with minimized reconstruction error is of interest.

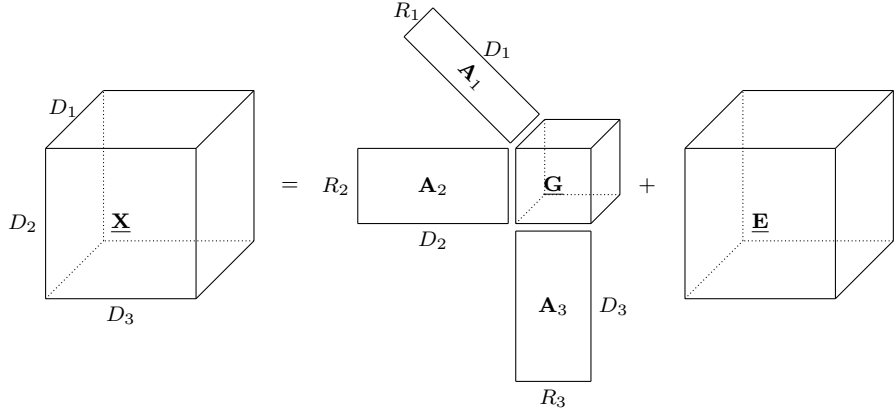
\begin{figure*}[t]
	\centering
	\input{hoda_fw.tikz.tex}
	\caption[A forward projection for \ac{hoda}.]{The forward projection for HODA.
		By calculating activation patterns $\mat{A}_k$, the original tensor $\ten{X}$ can approximately be
		reconstructed from projected latent tensor $\ten{G}$.
		The reconstruction is accurate up to an error term $\ten{E}$.
		$\mat{A}_k$ are chosen such that the variability captured in the latent tensor is
		maximally explained by the reconstructed tensor $\hat{\ten{X}}$ and not by
		the error term $\ten{E}$.}
	\label{fig:hoda-forward}
\end{figure*}

A straightforward and computationally efficient candidate for the \ac{hoda}
forward model, visualized in \cref{fig:hoda-forward}, is given as:
\begin{equation}
	\ten{X} = \ten{G}\mmpr{\mat{A}^\intercal} + \ten{E} =
	\hat{\ten{X}} + \ten{E}
	\label{eq:HODA-forward}
\end{equation}
with \emph{activation patterns} $\mat{A}_k \in \mathbb{R}^{D_k\times R_k}$,
reconstructed tensor $\hat{\ten{X}}$, and error term $\ten{E}$.

A good forward model should ensure that the norm of the reconstruction error
$\left\|\ten{E}\right\|_F$ is minimized.
In other words, variation captured in the latent tensor should be maximally captured by the
reconstruction term $\hat{\ten{X}}= \ten{G}\mmpr{\mat{A}^\intercal}$, and not by the error term
$\ten{E}$~\cite{Haufe2014}.
Hence, we aim to minimize the expected value of the cross-covariance between
the noise term and the extracted latent tensors:
\begin{equation}
	\left\{\mat{A}^*\right\}
	= \argmin_{\{\mat{A}\}}\text{E}\left[
		\text{vec}\left(\ten{E}(n)\right)\text{vec}\left(\ten{G}(n)\right)
		\right]_n
\end{equation}
or, equivalently~\cite{Parra2005,Haufe2014},
\begin{align}
	\left\{\mat{A}^*\right\}
	 & = \argmin_{\{\mat{A}\}}\sum_n^N\left[\ten{X}(n) -
	\hat{\ten{X}}(n)\right]^2                                                              \\
	 & = \argmin_{\{\mat{A}\}}\sum_n^N\left[\ten{X}(n) - \ten{G}(n)\mmpr{\mat{A}}\right]^2
\end{align}
This least-squares tensor approximation problem can be solved using the
Alternating Least Squares algorithm~\cite{Bentbib2022}, iteratively fixing all
but one of the activation patterns such that:
\begin{equation}
	\mat{A}_k = \argmin_{\mat{A}_k}
	\sum_n^N\left[\mat{X}_k(n) -
		\mat{A}_k\left(\ten{G}(n)\mmprs{\mat{A}}{k}\right)_k\right]^2
\end{equation}
at every iteration, which can be solved directly using ordinary least squares.
The activation patterns are initialized to the weights $\{\mat{U}\}$ of the
backward model.
Similar to fitting the backward model, the iterative process for the forward
model halts after a fixed number of iterations $I_\text{max}$ or when the update of each
$\mat{A}_k$ is lower than a predetermined threshold $\epsilon$.
The full procedure to determine the HODA forward projection is listed
in \cref{alg:HODA-fw}.
\begin{algorithm}
	\caption[A \acs{hoda} forward solution.]{The \acs{hoda} forward solution.}
	\label{alg:HODA-fw}
	\input{alg_hoda_fw.tex}
\end{algorithm}

\subsection{\Acl{bttda}}
After defining the forward model, we can construct the proposed block-term
tensor model.
Assuming the latent tensors $\ten{G}$ obtained by the backward projection of
HODA do not achieve perfect
class separation, the error term $\ten{E}$ in \cref{eq:HODA-forward} contains
some discriminative information.
This, in turn, can be exploited to improve classifier
performance.
Useful features can then be extracted from $\ten{E} = \ten{X} -
	\hat{\ten{X}}$ by further projecting it onto another core tensor
$\ten{G}^{(2)}$, assuming $\ten{G}$ as $\ten{G}^{(1)}$.

We thus extend the \ac{hoda} feature extraction scheme to \acf{bttda}.
\Ac{bttda} finds multiple discriminative blocks, such that its forward
model adheres to the block-term tensor structure:
\begin{equation}
	\ten{X} = \sum_b^B\ten{G}^{(b)}\mmpr{\mat{A}^{(b)}} + \ten{E}
	\label{eq:BTTDA-forward}
\end{equation}
for $B$ extracted latent tensors $\ten{G}^{(b)}$ and residual error term
$\ten{E}$.
The \ac{bttda} model is further illustrated by~\cref{fig:BTTDA}.
\begin{figure*}[t]
	\centering
	\input{bttda_fw.tikz.tex}
	\caption[A forward model for \acs{bttda}.]{A forward model for \acf{bttda}.
		\Ac{bttda} can extract more features
		than \ac{hoda} by iteratively finding a latent tensor $\ten{G}^{(b)}$ in a
		deflation scheme.
		The \ac{hoda} backward projection is first applied. Next, the
		input data is reconstructed via the HODA forward model and the
		difference between the two is found.
		Finally, this process is repeated with this difference as input data, until a
		desired number of blocks $B$ has been found.}
	\label{fig:BTTDA}
\end{figure*}
The block-term structure of this model implies that it is a generalization of both
the Tucker-structured \ac{hoda} and PARAFAC-structured discriminant feature
extraction.
If $B$ in \cref{eq:BTTDA-forward} is set to one, \ac{bttda} is equivalent to
\ac{hoda}; if at each term $b$ the dimension of the core tensor are
$(R_1^{(b)}=R_2^{(b)}=\ldots=R_k^{(b)}=1)$, a \ac{parafac} structure is assumed and
the resulting discriminant model is titled \ac{parafacda}.

Since \ac{bttda} is specified above as a forward model, a backward procedure
is required which finds the latent tensors $\ten{G}^{(b)}$ given $\ten{X}$ to
\ac{bttda} for feature extraction.
The extracted features represented by the latent tensors $\ten{G}^{(b)}$ can be
computed through a deflation scheme summarized in \cref{alg:BTTDA}.
\begin{algorithm}
	\caption{\Ac{bttda} feature extraction.}
	\label{alg:BTTDA}
	\input{alg_bttda.tex}
\end{algorithm}
For each block $b$, the latent tensor is extracted using the HODA backward
projection from the residual error term of the previous
block $\ten{E}^{(b-1)}$ as in \cref{eq:HODA-backward}:
\begin{equation}
	\ten{G}^{(b)} = \ten{E}^{(b-1)}\mmpr{\mat{U}^{(b)}}
\end{equation}
This residual error term is calculated by finding the difference between the
previous error and its reconstruction after backward and forward \ac{hoda}
projection:
\begin{align}
	\ten{E}^{(b)}
	 & = \ten{E}^{(b-1)} - \hat{\ten{E}}^{(b-1)}                      \\
	 & = \ten{E}^{(b-1)} - \ten{G}^{(b)}\mmpr{\mat{A}^{\intercal(b)}}
\end{align}
with $\ten{E}^{(0)}=\ten{X}$.

The resulting latent tensors can be vectorized and concatenated into
one single feature vector per input tensor:
\begin{equation}
	\mat{g}
	=\left[\vect\left(\ten{G}^{(1)}\right)\
		\vect\left(\ten{G}^{(2)}\right)\
		\cdots\
		\vect\left(\ten{G}^{(B)}\right)\right]
\end{equation}
so that they can be classified in a similar manner to HODA.

\subsection{Model and feature selection}
Similar to the unsupervised \ac{btd}, the performance of
\ac{bttda} is heavily dependent on the number of blocks $B$ and their
corresponding dimensions $\{(R_1^{(b)}, R_2^{(b)}, \ldots,	R_K^{(b)})\}_b^B$.
If these are not known a priori or can not set based on insights into the
data generation process, a model selection step is necessary in order to
determine the optimal values for $R_k^{(b)}$ and $B$.
These hyperparameters can be set through cross-validated hyperparameter tuning,
although computationally expensive.

To reduce the hyperparameter search space, we introduce
a single hyperparameter $\smash{\theta \in [0,1]}$ which replaces the block
dimensions $\{(R_1^{(b)}, R_2^{(b)}, \ldots,	R_K^{(b)})\}_b^B$.
The new hyperparameter $\theta$ then controls the sparsity of the \ac{bttda} solution, with $\theta=0$
corresponding to the \ac{parafacda} model with blocks of dimension $(1,1,\ldots,1)$, and $\theta=1$
corresponding to blocks of full rank $(D_1, D_2,\ldots,D_K)$.
For $0 < \theta < 1$, the dimension of block $b$ can be determined
analogous to the method described by \textcite{Phan2010}.
Here, $R_k$ are chosen based on the number of components needed to explain a
certain proportion of the variability in a mode of the input data for a
Tucker-structured decomposition.
For the \ac{hoda} model used in \ac{bttda}, this can be achieved using the eigenvalues of the per-mode total scatter matrix of tensor $\ten{E}^{(b-1)}$
\begin{equation}
	\mat{S}_{k,\text{t}}^{(b)} = \sum_n^N\mat{E}_k^{(b-1)}(n)\cdot\mat{E}_k^{(b-1)\intercal}(n)
	= \mat{W}_k^{(b)}\mat{\Lambda}_k^{(b)}\mat{W}_k^{(b)\intercal}
\end{equation}
such that
\begin{equation}
	R_k^{(b)} = \argmin_{R\in 1,\ldots,D_k}\frac{\sum_r^R\lambda_{k,r}^{(b)}}{\sum_r^{D_k}\lambda_{k,r}^{(b)}} > \theta
\end{equation}

Finally, \ac{hoda}, and by extension \ac{bttda}, can extract a substantial amount
of redundant features.
These should be dropped after projection and before proceeding to the classification
step~\cite{Phan2010}.
In \ac{bttda} in particular, redundant features can accumulate over the number of
blocks, hampering performance.
Furthermore, discriminant features across blocks can be heavily correlated since
all blocks are independently optimizing the same discriminability criterion.

To tackle these issues, extracted features are first decorrelated and scaled using
a whitening \ac{pca} transformation, retaining all principal components.
Relevant \ac{pca} components can be identified by calculating the
univariate Fisher score $\phi(i)$ for each component $i$ after \ac{pca},
calculated as
\begin{equation}
	\phi(i) = \frac
	{\sum_c^C N_c \left[\bar{g}_i(c)-\bar{\bar{g}}_i\right]^2}
	{\sum_n^N \left[g_i(n)-\bar{g}_i(c_n)\right]^2}
\end{equation}
Only features where $\phi(i) > 1$, i.e., between-class variance is greater
than within-class variance, are retained.
If there  are no extracted features with $\phi(i) > 1$, only the feature with the highest
$\phi(i)$ is retained.

\section{Experiments}
\subsection{Datasets and decoders}
We evaluated our proposed model in two offline \ac{eeg}-based \ac{bci} decoding problems:
the \acf{erp} and \acf{mi} paradigms using the openly available \ac{moabb} datasets
(version 1.2.0)~\cite{Aristimunha2023}.
\Ac{moabb} is widely accepted as a suitable benchmark for decoders aimed at
classical \ac{bci} problems, allowing fair comparison of machine learning classifiers
independent from data preprocessing.
Details about these datasets are found in \cref{tab:moabb}.
The \ac{erp} decoding task focuses on distinguishing target from non-target \acp{erp},
while the \ac{mi} tasks consists of distinguishing different imagined or performed
limb movements.
Within-session classification performance was assessed using stratified 5-fold
cross-validation. Performance was calculated as the \ac{rocauc} for binary
classification problems and accuracy for multi-class problems, in line with
\ac{moabb} benchmarking framework.
Average performance scores are balanced over dataset by taking the mean of
the per-dataset average performance scores.

To use \ac{hoda}, \ac{bttda}, and \ac{parafacda} as a decoder, they are paired
with \ac{lda} to classify the extracted features (HODA+LDA).
Hyperparameters candidates $\theta \in \left\{0, 0.1, 0.2, \ldots 1\right\}$
for all three decoders and $b \in\left\{1,2\ldots,16\right\}$ in the case
PARAFACDA+LDA and BTTDA+LDA
were tuned each evaluation fold using nested, stratified 5-fold cross-validation.
Other \ac{hoda} hyperparameters were set to $\epsilon=\num{1e-6}$ and $I_\text{max}=128$.

Differences in classification score between these proposed decoders
were statistically verified using one-sided Wilcoxon rank-sum tests performed per
dataset and decoder
pair on the cross-validated scores per subject and session.
Following the \ac{moabb} evaluation framework, meta-analyses for all \ac{erp} and \ac{mi} datasets respectively
were performed using the Stouffer method and effect size was determined as the
\ac{smd} between classification scores.

As additional comparison with other commonly used decoders, we selected a subset
of the decoders evaluated by \textcite{Chevallier2024}.
These decoders have been thoroughly evaluated on the \ac{moabb} benchmark to
identify them as generally accepted state-of-the-art methods.
For the \ac{erp} task, these included the Riemannian Geometry-based
classifiers ERPCov+MDM, ERPCovSVD+MDM, XDAWNCov+MDM, XDAWNCov+TS+SVM and the linear
classifier.
For the \ac{mi} task, the comparison methods were selected from Riemannian
methods ACM+TS+SVM, FgMDM, TS+EL, and the deep learning classifiers EEGTCNet
and ShallowConvNet.
We refer to \textcite{Chevallier2024} for the description, implementation details
and references of these methods.

\subsection{Event-Related Potentials}
\Acp{erp} are spatiotemporal features, with each sample forming a
second-order tensor with $K=2$ modes (a matrix), representing \ac{eeg}
channels and time samples
per epoch.

The \ac{erp} datasets listed in \cref{tab:moabb}
are first processed according to the \ac{moabb} framework.
\Ac{eeg} signals were recorded at the sample rate given
by \cref{tab:moabb} and band-pass filtered between 1 Hz
and 24 Hz.
The signals were cut into epochs starting from stimulus onset with a
dataset-specific length given by \cref{tab:moabb}.
For HODA+LDA, PARAFACDA+LDA, and BTTDA+LDA decoders, epochs were further
downsampled to 48 Hz.

\begin{figure*}[t]
	\input{pairwise_erp.tikz.tex}
	\vskip-1em

	\hskip-2em\input{pairwise_mi.tikz.tex}
	\caption{%
		Meta-analysis of decoder classification performance comparisons per dataset.
		Analyses were performed on \ac{rocauc} score for \ac{erp} datasets (top) and
		accuracy for \ac{mi} datasets (bottom).
		For the evaluated \ac{erp} datasets, \ac{bttda} always outperforms \ac{hoda}.
		\Ac{bttda} outperforms \ac{hoda} for 3 out of 5 \ac{mi} datasets.
		$***$: $p<0.001$; $**$: $p<0.01$, $*$: $p<0.05$.
	}
	\label{fig:results/meta}
\end{figure*}
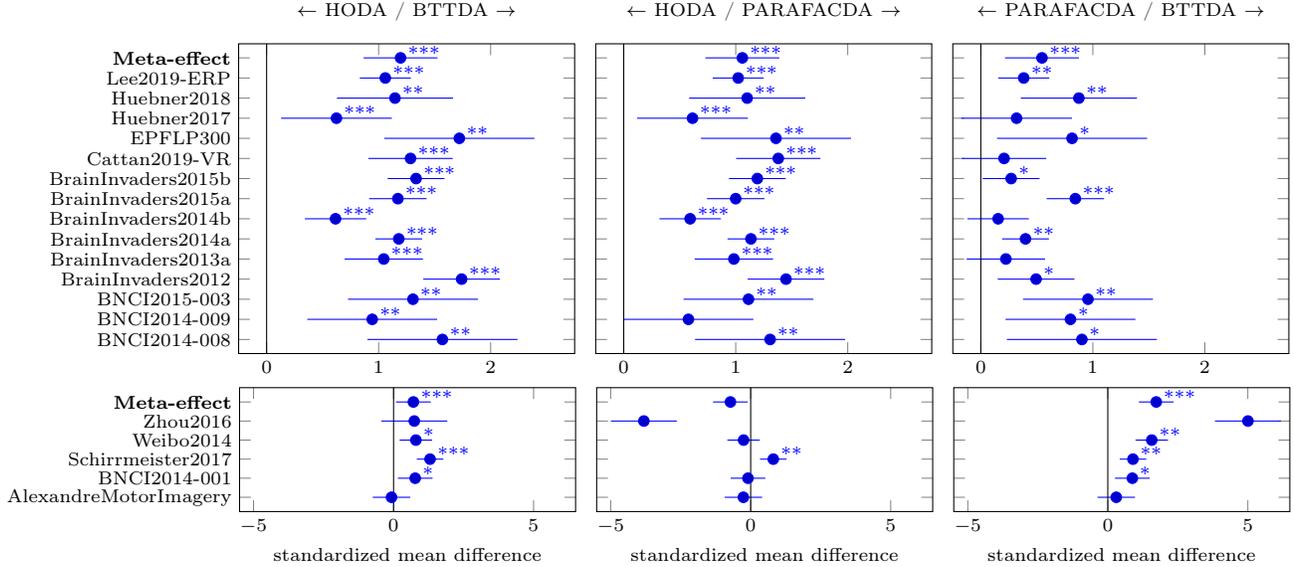

When considering grand average \ac{rocauc} over all evaluated \ac{erp} datasets
as reported in \cref{tab:results/erp/score},
the full BTTDA+LDA model (avg. \ac{rocauc}: 91.25$\pm$6.77\%) outperforms PARAFAC+LDA
(90.94$\pm$6.90\%),
and both in turn outperform HODA+LDA (88.89$\pm$7.04\%).
The meta-analysis shown in \cref{fig:results/meta} revealed the following
significant effects:
BTTDA+LDA > HODA+LDA ($p=\num{5.65e-65}$, SMD=$1.17$),
PARAFACDA+LDA > HODA+LDA ($p=\num{2.47e-58}$, SMD=$1.06$), and
BTTDA+LDA > PARAFAC+LDA ($p=\num{4.90e-15}$, SMD=$0.50$).
Both BTTDA+LDA and PARAFACDA+LDA always significantly outperform HODA+LDA.
BTTDA+LDA significantly outperforms PARAFACDA+LDA in 9 out of 14 datasets.
Significance and effect sizes for all evaluated \ac{erp} datasets are reported in~\cref{tab:results/erp/stats}.

\begin{sidewaystable*}
	\footnotesize
	\input{score_erp.tex}

	\caption{Area under the receiver operating characteristic curve for
		cross-validated within-session evaluation of HODA+LDA and our proposed decoders
		PARAFACDA+LDA and BTTDA+LDA evaluated on \ac{erp} datasets.
		Scores for other decoders were taken from \textcite{Chevallier2024}.
		BTTDA+LDA always outperforms HODA+LDA and PARAFACDA+LDA, except for datasets,
		and consistently is nearly on par with or outperforms
		the state-of-the-art XDAWNCov+TS+SVM decoder.
	}
	\label{tab:results/erp/score}
\end{sidewaystable*}
Compared to the state-of-the-art XDAWNCov+TS+SVM decoder, BTTDA+LDA scores
better in 8 out of 14 datasets, combined with a moderate increase in grand average \ac{rocauc}
($91.25 > 90.82$).

Full cross-validation results can be retrieved from additional file \cref{item:add/mi-results}.

\subsection{Motor Imagery}

For \ac{mi}, discriminatory information is encoded in the \ac{eeg} data as
\acp{ersd}.
Contrary to the time-domain analyses performed on \acp{erp}, \acp{ersd} are
discerned in the power expressed in the time-frequency domain.
For the \ac{mi} task, we transform the \ac{eeg} signal into the
time-frequency domain, forming third-order tensors, with $K=3$ modes
respectively representing channels, frequencies, and time bins.

To achieve this, the \ac{eeg} signals in the \ac{mi} datasets listed in \cref{tab:moabb}
are first processed using the \ac{moabb} motor imagery pipeline.
\ac{eeg} signals were recorded at the sample rate given
by \cref{tab:moabb} and band-pass filtered between 8 Hz
and 32 Hz.
The signals were then cut into epochs starting from stimulus onset with a
dataset-specific length given by \cref{tab:moabb}.
Custom postprocessing to convert epochs to third-order tensors extracted
the magnitude of the complex Morlet-wavelet transform with 17 logarithmically spaced frequencies from 8 Hz to 32 Hz and a varying number of cycles logarithmically spaced from 4 to 16.
Finally, the magnitude envelope was downsampled to 32 Hz using an anti-aliasing
filter and decimation.

In line with the \ac{moabb} method, only the first three classes per dataset were
used.
When considering grand average classification accuracies over all evaluated
\ac{mi} datasets as reported in \cref{tab:mi-score},
the full BTTDA+LDA model (avg. accuracy: $64.52\pm12.23$\%)
outperforms PARAFACDA+LDA ($58.89\pm11.27$\%) and HODA+LDA
($61.00\pm11.11$\%).
The meta-analysis shown in \cref{fig:results/meta} revealed the following significant effects:
BTTDA+LDA > HODA+LDA ($p=\num{6.20e-5}$, SMD=$0.75$),
BTTDA+LDA > PARAFAC+LDA ($p=\num{4.00e-6}$, SMD=$1.48$).
BTTDA+LDA outperforms HODA+LDA except for datasets Zhou2016 and AlexandreMotorImagery.
PARAFACDA+LDA outperforms HODA+LDA for dataset Schirrmeister2017.
BTTDA+LDA outperforms PARAFACDA+LDA except for datasets Zhou2016 and AlexandreMotorImagery.
Significance and effect sizes for all evaluated \ac{mi} datasets are reported in~\cref{tab:results/mi/stats}.

\begin{sidewaystable*}
	\footnotesize
	\input{score_mi.tex}

	\caption{Cross-validated classification accuracies for within-session evaluation
		to
		of HODA+LDA and our proposed decoders	PARAFACDA+LDA and BTTDA+LDA,
		evaluated on three-class motor imagery datasets.
		Tensor-based methods generally score lower than Riemannian Geometry-based
		decoders.
		\Ac{bttda} outperforms
		Accuracies for other decoders were taken from \textcite{Chevallier2024}.}%
	\label{tab:mi-score}%
\end{sidewaystable*}
All of HODA+LDA (avg. accuracy $61.00\pm11.11$) and our proposed decoders PARAFACDA+LDA
($58.89\pm11.27$) and BTTDA+LDA ($64.52\pm12.23$) score
substantially lower than state-of-the-art decoder ACM+TS+SVM ($75.77\pm11.12$).

Full cross-validation results can be retrieved from additional file \cref{item:add/mi-results}.

\subsection{Impact of block dimension and number of blocks}

\begin{figure}[ht]
	\footnotesize
	\input{gridsearch.tikz.tex}
	\caption{%
		Cross-validated BTTDA+LDA \ac{rocauc} (left) and \ac{bttda} \ac{nmse}
		(right) for dataset BNCI2014-008 as a function of the number of blocks $b$
		and the hyperparameter $\theta$	which controls the block dimensions.
		More effective class separation occurs as $b$ increases while \ac{nmse}
		decreases. Eventually, overfitting occurs and class separation performance
		drops or plateaus depending on the effectiveness of feature selection as
		shown here.
	}
	\label{fig:blocks}
\end{figure}
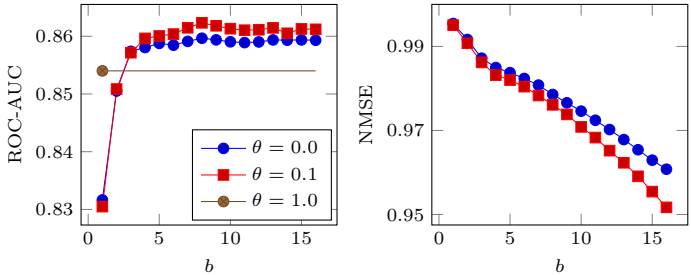

To analyze the contribution of extra feature blocks extracted by BTTDA over
the first one found by HODA, we perform the following analyses on \ac{erp} dataset
BNCI2014-008 chosen for its minimal computational requirements.
We investigated cross-validated within-session \ac{rocauc} scores as function
of the number of blocks ($b$) and hyperparameter $\theta$, shown in \cref{fig:blocks} (left)
averaged over all subjects.
$b$ was varied from 1 to 16, while $\theta$ was chosen from
$\smash{\left\{0.0, 0.1, 0.2,\ldots, 1.0\right\}}$.
Full results are presented in additional file \cref{item:add/blocks}.
Below, we report on selected of $\theta$ choices.
When $\theta=0$, the \ac{bttda} model corresponds to the \ac{parafacda} decoder.
$\theta=0.1$ yielded the highest BTTDA+LDA \ac{rocauc}.
$\theta=1$ resulted in the highest overall HODA+LDA ($b=1$) performance.
At $\theta=1$, no blocks other than the initial block can be modeled, since
$\theta=1$ by definition explains all data in the dataset and further forward
modeling fails.

At $b=1$, corresponding to the \ac{hoda} model, sparse models with $\theta=0$ (avg.
\ac{rocauc} 83.16\%) and
$\theta=0.1$ (avg. \ac{rocauc} 83.05\%) are substantially lower than the
optimal performance at $\theta=1.0$ (avg. \ac{rocauc} 85.40).
Moving from the \ac{hoda} model ($b=1$) to the \ac{bttda} and \ac{parafacda}
model allows the extraction of more blocks ($b\geq1$).
With this relaxation, \ac{parafacda} and \ac{bttda} ($\theta=0.1$) exceed HODA
the at $b=3$ (avg. \ac{rocauc} 85.74\% and 85.71\% respectively), while
maintaining lower reduced dimensions than the (high) optimal dimensions for
\ac{hoda} ($\theta=1.0$).
Eventually, \ac{bttda} its reaches the highest overall \ac{rocauc} at $b=8$
(avg. \ac{rocauc} 86.23\%).
In general, when only a single block is used, a high $\theta$ is needed.
When more blocks are used, higher $\theta$ deteriorates performance.
Higher performance can be reached by choosing a low $\theta$ and $b>1$, resulting
in multiple blocks with low dimensions.

Additionally, \cref{fig:blocks} (right) shows the effectiveness of the forward
modeling step measured as the cross-validated \ac{nmse} when reconstructing
the original data  from the truncated \ac{bttda} decomposition
$\textstyle{\hat{\ten{X}}^{(B)}=\sum_b^B\ten{G}^{(b)}\mmpr{\mat{U}^{(b)}}}$,
with \ac{nmse} is calculated as:
\begin{equation}
	\nmse\left(\ten{X}, \hat{\ten{X}}^{(B)}\right) =
	\frac{\sum_n^N\left\|\ten{X}(n)-\ten{\hat{X}}^{(B)}(n)\right\|_\text{F}^2}
	{\sum_n^N\left\|\ten{X}(n)\right\|_\text{F}^2}
\end{equation}

\Ac{nmse} decreases monotonically with $b$ for both $\theta=0.0$ and $\theta=0.1$.
In general, \ac{nmse} decreases faster as $\theta$ increases.
For $\theta=1$, reconstruction \ac{nmse} at $b=0$ is near zero ($\num{1.32e-30}$)
since no information is lost in the full-rank decomposition.

\subsection{Interpretable decomposition}

The following qualitative analysis reveals the model interpretability provided
by the forward modeling step, by relating patterns in the reconstructed data
to expected effects visible in the neural data at hand.

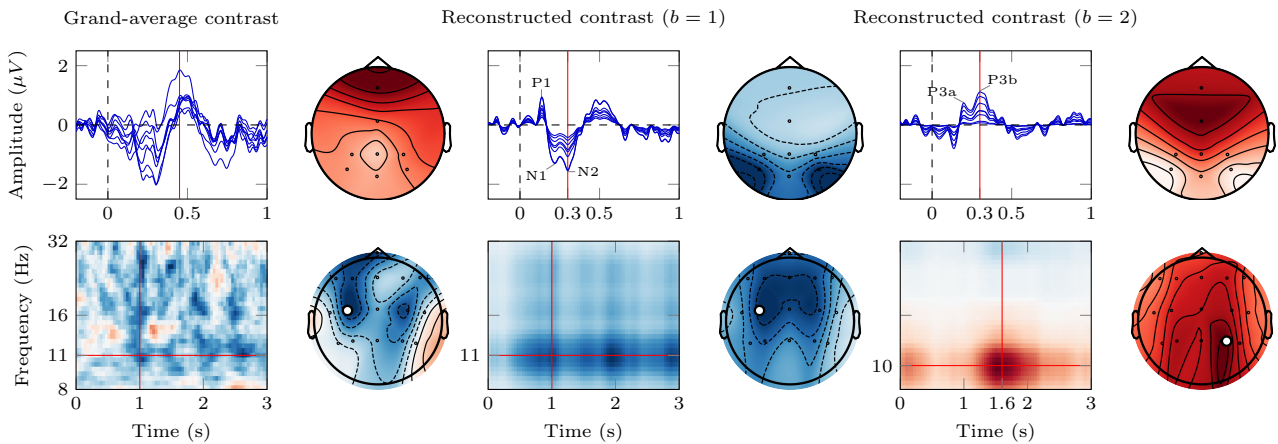
\begin{figure*}
	\input{erp_contrast.tikz.tex}

	\input{mi_contrast.tikz.tex}

	\caption{%
		Per-block forward \ac{bttda} model activation pattern contrasts and
		overall grand-average contrast for \ac{erp} dataset BNCI2014-008 (top) and
		\ac{mi} dataset AlexMI (bottom).
		Red lines indicate the slices generating the scalp plot.
		In the bottom row, the white dot indicates for which channel the time-frequency
		spectrum was plotted.
		The \ac{erp} is decomposed in parieto-occipital components (P1, N1, N2)
		corresponding to visual processing and fronto-central components (P3a,P3b)
		related to task processing.
		The right-hand \acf{mi} \ac{ersd} is decomposed in mostly contralateral high-$\mu$ band
		desynchronization, and parietal $\alpha$-band	synchronization.
	}
	\label{fig:results/interpret}
\end{figure*}

All three proposed models were trained on the combined subjects in BNCI2014-008
for \ac{erp} classification and AlexMI for \ac{mi}.
To allow proper visual inspection, the \ac{erp} epochs were extended with
a pre-stimulus interval of 0.2 s for baseline correction and the original sample
rate of 256 Hz was kept.
The \ac{mi} epochs were sampled at 250 Hz after time-frequency transformation.
The number of blocks in this example was set to $B=2$ and hyperparameters $\theta$ were tuned
using 5-fold stratified cross-validation with entire-subject holdouts to
determine the best hyperparameter for cross-subject decoding
(\ac{erp}: $\theta=0.3$, \ac{mi}: $\theta=0.7$).
Each model was retrained with these hyperparameters on the full data combined
over all subjects.
Using these models,
These models then generated reconstructed contrasts
$\bar{\ten{C}}_{c_2-c_2}^{(b)}$ between classes $c_2$ and $c_1$ for $b=1$ and
$b=2$ as in \begin{equation}
	\bar{\ten{C}}_{c_2-c_1}^{(b)} = \left[
		\bar{\ten{G}}_{c_2}^{(b)}
		- \bar{\ten{G}}_{c_1}^{(b)}
		\right]\mmpr{\mat{A}^{(b)}}
\end{equation}
with $c_2$ target and $c_1$ non-target trials for \ac{erp}, and $c_2$ right hand
imagery and $c_1$ rest for \ac{mi}.
These contrasts, together with the grand-average contrast, are shown in
\cref{fig:results/interpret}

The grand-average \ac{erp} contrast shows an entangled superposition
of several different \ac{erp} components~\cite{Luck2011}.
The activation patterns of the first two blocks disentangle this contrast
in effects that can be related to \ac{erp} literature in the context of the
classic visual P300 matrix speller task in BNCI2014-008~\cite{Riccio2013}.

Block 1 exhibits positive and negative peaks in the lateral parieto-occipital
regions corresponding to the visual cortex.
The first positive peak and 2 negative peaks (P1, N1, N2)
correspond to early components reflecting the task-related visual processing
modulated by a mix of visual fixation and visual attention~\cite{Treder2010}.
Block 2 has a more central scalp expression, and shows 2 positive peaks (P3a, P3b).
Together with the residual positive activation between 0.4 s and 0.6 s in block
1, these constitute the processing of the attention-modulated detection of rare
stimuli present in the P300 matrix speller task~\cite{Kamp2013}.

For motor imagery, results are displayed in the time-frequency domain.
Positive values indicate event-related synchronization, negative values
desynchronization.
Upon visual inspection, the grand-average contrast shows no dominant pattern of
synchronization or desynchronization, possibly due to the limited dataset size.

\Ac{bttda} decomposition extracts two distinct effects.
Block 1 shows a persistent desynchronization between 9Hz and 13Hz most prominent
in the left central area.
For right-hand motor imagery, this corresponds to expected task-specific and
localized high-$\mu$ band desynchronization in the contralateral motor
cortex~\cite{Pfurtscheller2000,Wolpaw2012}.
Block 2 exhibits a synchronization between 8 and 12 Hz over the
parieto-occipital region, from 1.2 to 2.2 s.
This may be interpreted as the $\alpha$ band surround-ERS observed during hand
movement~\cite{Suffczynski1999, Gerloff1998,Wolpaw2012}.

\section{Discussion}
\subsection{Contribution}

The \ac{hoda} model used for \ac{bci} decoding can be constrained by its
Tucker structure.
We introduced a more flexible generalization termed \ac{bttda} with a
block-term tensor structure.
We also introduce \ac{parafacda}, a special case of \ac{bttda} expressed as a
sum of multilinear rank-1 terms.
Our results show that \ac{bttda} consistently scores on par or significantly higher than
\ac{hoda} as a supervised dimensionality reduction technique for \ac{bci} decoding.
\Ac{bttda} managed to outperform \ac{hoda} with 2.36\%pt. on average for
\ac{erp} datasets, and 2.75\%pt. for \ac{mi} datasets.
\Ac{parafacda} also scored 0.31\%pt. higher than  \ac{hoda} in \ac{erp} datasets
but was outperformed by \ac{bttda} overall.

BTTDA yields state-of-the-art decoding performance for \ac{erp}
datasets in the \ac{moabb} benchmark, but fails to do so for \ac{mi} datasets.
While this effect is not consistent over all \ac{erp} datasets, and the
increase is often rather low (<2\%pt.), it averages out over all datasets
as a moderate increase of 0.43\%pt.
We note that performances of other decoders for these problems already achieve
relatively high binary classification performance, which does not always leave
room for improvement.

As mentioned above, results for \ac{mi} were substantially lower than expected.
Not only does \ac{bttda} perform poorly, but the baseline \ac{hoda} model as well.
This is unexpected since it conflicts with literature which uses \ac{hoda} to
effectively classify \ac{mi} from time-frequency transforms~\cite{Phan2010,Lotte2018,Liu2015,Cai2021}.
If the issues hampering \ac{hoda} performance can be identified, \ac{bttda}
could gain ground on the state of the art.
We believe poor \ac{mi} performance in our case could stem from the following issues.
The time-frequency decomposition and data transformations or their parameters
used in this study might not be suited to capture the relevant \ac{ersd}
information necessary for performant classification.
For fair comparison with other MOABB decoders, the standard MOABB preprocessing
pipeline was followed, which might interfere with our postprocessing.
In this case, \ac{mi} decoding might benefit from different preprocessing,
transformation or	tensorization techniques.
On the other hand, hyperparameter selection could require more candidates or
cross-validation folds due to the combination of $K=3$ and larger data size following to the
time-frequency transformation.
Solutions for this problem can be computationally expensive.

Finally, due its the inherent forward modeling
steps, \ac{bttda} is intrinsically an explainable model which allows for interpretation
of the signal components modeled by the tensor blocks.
While the weights of the backward projections are
hard to interpret~\cite{Haufe2014}, the activation patterns and contrasts after
forward projection can reveal patterns in neural data.
Qualitative analyses showed that block activation patterns correspond to
task-related physiological processes for both \ac{erp} and \ac{mi}
classification problems.
Given informed or correctly tuned hyperparameters, this method could be used to,
e.g., separate and identify neural processes based on the task-related
information in the class labels.
More generally, \ac{bttda} can achieve an effective unmixing of signal generators
relevant to the classification problem at hand, which might otherwise not be
properly separated within the constraints of the \ac{hoda} model.
A point of care, however, arises from the deflation scheme: some processes
might already be partially explained by previous blocks.
In this case, information from a single physiologic process might not be modeled
using only a single block.
Hence, previous blocks might need to be taken into account to properly interpret
a block activation pattern.

\subsection{Modeling assumptions}

We assume the main benefit of \ac{bttda} stems from the following two aspects.
Given fixed block dimensions, extra \ac{bttda} blocks with proper feature selection
can discover more discriminant information over \ac{hoda}.
While no proof is given here, we show that NMSE monotonically decreases.
This suggests that all the variation in the signal will eventually be explained
by the model while still extracting features that are maximally discriminant.
Eventually, the number of blocks will reach a point of diminishing validation
score returns.
At this point, adding extra features to the decision classifier increases
the risk of overfitting instead of adding extra useful discriminatory
information.
Hence, performance increases with the number of blocks until overfitting occurs.

On the other hand, a \ac{bttda} solution is more parsimonious than a \ac{hoda}
solution can achieve due to its block-term structure compared to \ac{hoda}'s full Tucker
structure, as illustrated	by \cref{fig:bttda/sparse}.
In other words, the same discriminative information captured by a relatively large
Tucker-structured core tensor could be expressed more sparsely with a small
number of block-terms, while avoiding redundant features.
The \ac{parafac} structure employed in \ac{parafacda} is even more sparse, which could be
a benefit or a drawback depending on the amount of regularization required,
or on the true underlying structure of the data.
\Ac{bttda} with a few, sparse blocks might perform worse then a dense \ac{hoda}
solution, adding extra \ac{bttda} blocks eventually overpasses the \ac{hoda}
solution as indicated by~\cref{fig:blocks}.

The enhanced performance could also partially stem from \ac{bttda}'s internal
model of the data covariance.
Since HODA estimates one within-class scatter matrix
$\mat{S}_{-k,\text{w}}\in\mathbb{R}^{D_k\times D_k}$ per mode during training,
its overall model of the data scatter is determined by these per-mode scatter matrices as a
Kronecker product $\mat{S}_{-1,\text{w}}\otimes \mat{S}_{-2,\text{w}}\otimes\cdots\otimes \mat{S}_{-K,\text{w}}$.
This corresponds to the assumption that the \ac{eeg} data is
drawn from a multilinear normal distribution~\cite{Ohlson2013}.
Similar assumptions are made in \ac{erp} decoding algorithms such as
Spatial-Temporal Discriminant-Analysis~\cite{Zhang2013} and LCMV-beamforming
with Kronecker covariance structure~\cite{Kerchove2022}.
However, it is known that \ac{eeg} covariance cannot fully be expressed as a
single Kronecker product.
Rather, it is more accurately modeled as a sum of
multiple Kronecker products~\cite{Bijma2005, Sosulski2022}.
Since \ac{bttda} iteratively fits \ac{hoda} models to the residual error, each with its own
multilinear covariance model, it allows modeling multiple different multilinear
covariance terms, refining the internal covariance model.
This way, multiple effects with corresponding multilinear distributions
can be extracted.

Finally, the drop in performance for \ac{parafacda} in \ac{mi} datasets is
attributed to the \ac{parafac} interaction of the model's rank-1 term structure
with the multi-class nature of the \ac{mi} problems.
\Ac{parafacda} only extracts a single feature per block, which cannot properly
separate more than 2 classes.
Further blocks are not properly adapted to take into account which classes
have been separated by earlier blocks, hence extracting more \ac{parafac}
blocks might not be helpful.

In summary, we conclude there is an effective	added value in iteratively extracting
multiple block terms.
The flexibility of the \ac{bttda} model is both expressed in its ability
to capture more discriminant information with more parsimony,
and in its ability to capture effects which cannot be expressed by the \ac{hoda}
model, such as the \ac{eeg} covariance structure.
This makes it specifically suited to tackle classification problems encountered in
brain-computer interfacing.

\subsection{Model selection}

\Ac{bttda} trades the rigid \ac{hoda} model for increased model complexity with more
hyperparameters to tune, which expands the solution space to settings where
performance can be improved.
Extracting more blocks and tuning the hyperparameters increases the time
complexity of fitting \ac{bttda}-based models compared to \ac{hoda}-based models.
\Ac{hoda} feature extraction can be solved with time complexity
\begin{equation}
	\mathcal{O}\left(\left|\Theta\right|FI_\text{max}NK^2D^{K+1}\right)
\end{equation}
whereas \ac{bttda} training and model selection increases this to
\begin{equation}
	\mathcal{O}\left(\left|\Theta\right|FBI_\text{max}NK^2D^{K+1}\right)
\end{equation}
with $\Theta$ the set of $\theta$ candidates and $F$ the number of
cross-validation folds for hyperparameter tuning.
\Cref{app:complexity-derivation} shows complexity derivation.
Overall, the \ac{bttda} approach shifts the focus of tensor discriminant analysis
from finding optimal projections to model selection driven by computation.

The proposed $\theta$-controlled selection procedure efficiently
reduces the computational demand compared to tuning all hyperparameters
$\textstyle{\left\{  \left(R_1^{(b)},R_2^{(b)},\ldots,R_K^{(b)}\right)\right\}_b^B}$.
On the other hand, it also limits the chosen dimensions of each block to lie
within a subset of all possible configurations.
In a sense, this goes against the earlier proposition of increased model
flexibility.
Instances could occur where \ac{bttda} offers little to no added value over the
Tucker-structured \ac{hoda} when both are given totally free choice of
dimensions, but cases where \ac{bttda} could achieve greater performance could
equally be found.
Finding these optimal-dimension configurations, can currently only be achieved
through a costly, cross-validated hyperparameter search jointly over the
dimensions of each block.
Applications such as light-weight or mobile brain-computer interfaces should
carefully weight potential performance gains against this computational demand.
Future efforts should focus on more advanced automated hyperparameter selection
methods relying on sparsity criteria, eigenvalue truncation or information
criteria such as the ones used in \ac{bttr}~\cite{Faes2022}, or other
statistical measures depending on the application of the model.

Finally, we note that our proposed model selection procedure does not
guarantee grouping coherent projections within the same block according to some
desirable metrics.
Features across blocks are heavily correlated, leading to a high
degree of multicollinearity.
Currently, this is corrected  post-hoc by applying whitening and PCA.
Solutions imposing some sense of subspace orthogonality between the extracted
blocks could lead to a more effective feature extraction solution.
Sparsity, pattern interpretability, minimal or maximal within-block feature
correlation and ordering of blocks by decreasing discriminability are all
examples of useful within-block grouping criteria.

As future work, The impact of higher-order tensors ($K>3$) should be thoroughly
investigated, since this could have a large impact on model behavior.
We expect a dimensionality limit beyond which the forward modeling step cannot
accurately regress from the low-dimensional latent tensors to the
high-dimensional original tensors, introducing error in the input data for the
next block which can stack up over blocks.
The forward multilinear least squares problem is underdetermined hence prone to
numerical instability, which calls for a suited regularization approach.
Finally, other tensorization methods of the \ac{eeg} data should be explored,
such as time-lagged Hankel tensors~\cite{Papy2005} or tensors across subjects,
conditions or sliding windows if they are appropriate given the available prior
knowledge of the dataset.

\section{Conclusion}

We have introduced \acf{bttda}, a novel,
tensor-based, supervised dimensionality reduction technique optimized for class
discriminability, which adheres to the block-term tensor structure.
\Ac{bttda} is a generalization of \acf{hoda} and can also be
applied as a special sum-of-rank-one tensors \ac{parafacda} model.
The model is obtained by iteratively fitting \ac{hoda} in a deflation scheme,
leveraging a novel forward modeling step.

Via accompanying model selection hyperparameters, \ac{bci} decoders using
\ac{bttda} feature extraction can significantly outperform decoders based on
\ac{hoda} exceed state-of-the-art decoding performance on \acl{erp} problems
(second-order tensors) and outperform \ac{hoda} in motor imagery problems
(third-order tensors).
The inherent forward model of \ac{bttda} also allows interpreting the
discriminative processes considered by the classifier.

Moving from the rigid Tucker tensor structure of \ac{hoda} to the more flexible
and sparse block-term structure shifts the focus from finding the best constrained
multilinear projections to model and feature selection.
This approach allows performance and generalization to be traded for computational cost,
which is particularly relevant for \ac{bci} decoding problems.
Because of its general implementation and minimal assumptions on data structure,
\ac{bttda} can equally be applied to classification for other neuroimaging modalities
(MEG, ECoG, fNIRS, fMRI, EMG, etc.), or to tensor classification problems in other
domains.

\section*{Code availability}

The source code of the proposed \ac{bttda} algorithm and the analyses performed in
this work are available at \url{https://github.com/arnevdk/bttda}.

\section*{Additional data and materials}
\begin{enumerate}
	\item\textbf{Full ERP decoding cross-validation results} \\
	file: \texttt{erp\_results.csv}\\
	format: \textit{comma-separated values file}
	\label{item:add/erp-results}
	\item\textbf{Full MI decoding cross-validation results} \\
	file: \texttt{mi\_results.csv}\\
	format: \textit{comma-separated values file}
	\label{item:add/mi-results}
	\item\textbf{Full results of analysis in function of the number of blocks and block dimension}\\
	file: \texttt{block\_theta\_results.csv}\\
	format: \textit{comma-separated values file}
	\label{item:add/blocks}
\end{enumerate}

\section*{Acknowledgments}
We thank the Flemish Supercomputer Center (VSC) and the High-Performance
Computing (HPC) center of KU Leuven for allowing us to execute our
computational experiments on their systems.
We also wish to acknowledge Dr.\ Axel Faes for his inspiration in
conceptualizing this work.

AVDK has been is funded by the Belgian Fund for Scientific Research Flanders
(G0A4321N) and received support of the special research fund of the KU Leuven
(GPUDL/20/031) and the University of Lille under the Global PhD Scholarship Program.
MMVH is supported by research grants received from the European Union’s
Horizon Europe Marie Sklodowska-Curie Action program
(grant agreement No. 101118964), the European Union’s Horizon 2020 research and
innovation program (grant agreement No. 857375), the special research fund of
the KU Leuven (C24/18/098), the Belgian Fund for Scientific Research – Flanders
(G0A4118N, G0A4321N, G0C1522N), and the Hercules Foundation (AKUL 043).

The authors acknowledge the support of the RITMEA project co-financed by the
European Union with the European Regional Development Fund, the French state,
and the Hauts-de-France Region Council.

\printbibliography%
\clearpage%

\appendix
\setcounter{table}{0}
\renewcommand{\thetable}{\thesection\arabic{table}}

\include{complexity_derivation.tex}

\onecolumn

\section{Datasets}

\begin{table*}[!htbp]
	\footnotesize
	\input{moabb_datasets.tex}
	\caption{MOABB datasets used for evaluation, with the number of
		subjects (\# Sub.), the number of EEG channels (\# Chan.), the number of trials or trials per class for ERP
		datasets (\# Trials), the epoch length (Epoch len.), the sampling
		frequency (S. freq.), the number of sessions per subject (\# Sess.) and the
		number of runs (\# Runs). \ac{erp} datasets contain 2 classes, for \ac{mi} datasets the first 3 classes were retained. \Ac{erp} dataset Sosulski2019 was omitted due to technical problems.
		\Ac{mi} dataset PhysionetMI was omitted due to its high computational and
		storage demands.
		Adapted from~\cite{Aristimunha2023}
		and~\cite{Chevallier2024}.}%
	\label{tab:moabb}
\end{table*}

\newpage

\section{Pairwise statistics}
\begin{table*}[!htbp]
	\footnotesize
	\input{stats_erp.tex}
	\caption{Results of one-sided Wilcoxon rank-sum tests comparing the
		per-subject cross-validated classification scores of the evaluated \ac{erp}
		decoders.
		Significance is reported as $p$, the effect size as the standardized mean
		difference (SMD).}
	\label{tab:results/erp/stats}
\end{table*}

\begin{table*}[!htbp]
	\footnotesize
	\input{stats_mi.tex}
	\caption{Results of one-sided Wilcoxon rank-sum tests comparing the
		per-subject cross-validated classification scores of the evaluated \ac{mi}
		decoders.
		Significance is reported as $p$, the effect size as the standardized mean
		difference (SMD).}
	\label{tab:results/mi/stats}
\end{table*}

\end{document}

%% file: abstract.txt
Brain-computer interfaces (BCIs) allow direct communication between the brain and external devices, frequently using electroencephalography (EEG) to record neural activity. Dimensionality reduction and structured regularization are essential for effectively classifying task-related brain signals, including event-related potentials (ERPs) and motor imagery (MI) rhythms. Current tensor-based approaches, such as Tucker and PARAFAC decompositions, often lack the flexibility needed to fully capture the complexity of EEG data. This study introduces Block-Term Tensor Discriminant Analysis (BTTDA): a novel tensor-based and supervised feature extraction method designed to enhance classification accuracy by providing flexible multilinear dimensionality reduction. Extending Higher Order Discriminant Analysis (HODA), BTTDA uses a novel and interpretable forward model for HODA combined with a deflation scheme to iteratively extract discriminant block terms, improving feature representation for classification. BTTDA and a sum-of-rank-1-terms variant PARAFACDA were evaluated on publicly available ERP (second-order tensors) and MI (third-order tensors) EEG datasets from the MOABB benchmarking framework. Benchmarking revealed that BTTDA and PARAFACDA significantly outperform the traditional HODA method in ERP decoding, resulting in state-of-the art performance (ROC-AUC=91.25\%). For MI, decoding results of HODA, BTTDA and PARAFACDA were subpar, but BTTDA still significantly outperformed HODA (64.52\%>61.00\%). The block-term structure of BTTDA enables interpretable and more efficient dimensionality reduction without compromising discriminative power. This offers a promising and adaptable approach for feature extraction in BCI and broader neuroimaging applications.

%% file: tensor_core_structures.tikz.tex
\bigskip
\footnotesize
\begin{tikzpicture}[x=\textwidth/14.2, y=-\textwidth/14.2]
	\TensorThree{data}{}{}{}{2}{2}{2}
	\begin{scope}[shift={((5.5,0)}]
		\TensorThree{core}{}{}{}{1}{1}{1}
		\begin{scope}[shift={(-0.0707, -0.0707)}]
			\MatrixSkewed{fac. 1}{}{}{1}{1}
		\end{scope}
		\begin{scope}[shift={(-0.1,0)}]
			\MatrixLeft{fac. 2}{}{}{1}{1}
		\end{scope}
		\begin{scope}[shift={(0,1.1)}]
			\MatrixBelow{fac. 3}{}{}{1}{1}
		\end{scope}
	\end{scope}
	\begin{scope}[shift={(0,4)}]
		\TensorThree{$\ten{G}$}{}{}{}{2}{2}{2}
		\node[anchor=north, align=center] at (1,2) {Tucker structure};

		{\footnotesize
		\begin{scope}[shift={(5,0)}]
			\TensorThree{}{}{}{}{2}{2}{2}
			\node at (.33,.33,-.33) {$g^{(1)}$};
			\node at (.66,.66,-.66) {$g^{(2)}$};
			\node at (1,1,-1) {$\ddots$};
			\node at (1.5, 1.5, -1.5) {$g^{(b-1)}$};
			\node at (1.8, 1.9, -1.8) {$g^{(B)}$};
		\end{scope}
		}
		\node[anchor=north, align=center] at (6,2) {PARAFAC structure};

		\begin{scope}[shift={(10,0)}]
			\TensorThree{}{}{}{}{2}{2}{2}
			\TensorThree{$\ten{G}^{(1)}$}{}{}{}{.5}{.5}{.5}
			\node at (.75,.75,-.75) {$\ddots$};
			\begin{scope}[shift={(1,1,-1)}]
				\TensorThree{$\ten{G}^{(B)}$}{}{}{}{1}{1}{1}
			\end{scope}
		\end{scope}
		\node[anchor=north, align=center] at (11,2) {block-term structure};
	\end{scope}
	\draw[->] (2,1, -1) -- (3.9,1,-1);
	\draw (6.7,1,-1) -| (7.4,3, -1);
	\draw[->] (6, 3,-1) -| (1,4,-1);
	\draw[->] (6, 3,-1) -| (6,4,-1);
	\draw[->] (6, 3,-1) -| (11,4,-1);
\end{tikzpicture}

%% file: hoda_bw.tikz.tex
\begin{tikzpicture}[y=-1cm]
	\TensorThree{$\ten{G}$}{$R_1$}{$R_2$}{$R_3$}{1}{1}{1}
	\node at (2,0.5) {$=$};
	\begin{scope}[shift={(4,0)}]
		\TensorThree{$\ten{X}$}{}{}{}{2}{2}{2}
		\begin{scope}[shift={(-0.0707, -0.0707)}]
			\MatrixSkewed{$\mat{U}_1$}{$D_1$}{$R_1$}{1}{2}
		\end{scope}
		\begin{scope}[shift={(-0.1,0)}]
			\MatrixLeft{$\mat{U}_2$}{$D_2$}{$R_2$}{1}{2}
		\end{scope}
		\begin{scope}[shift={(0,2.1)}]
			\MatrixBelow{$\mat{U}_3$}{$D_3$}{$R_3$}{2}{1}
		\end{scope}
	\end{scope}
\end{tikzpicture}

%% file: alg_hoda_bw.tex
\begin{algorithmic}[1]
	\Require $\{\ten{X}(n)\}_n^N, \{c_n\}_n^N, (R_1,\ldots,R_K), I_\text{max}, \epsilon$
	\State $\mat{U}_k \gets $ orthonormal matrix $\in \mathbb{R}^{D_k\times R_k}
		\ \forall k$
	\State $\mat{S}_{k,\text{t}} \gets
		\textstyle{\sum_n^N}\mat{X}_k(n)\cdot\mat{X}_k^\intercal(n)\ \forall k$
	\State $i\gets 1$
	\Repeat
	\For{$k=1,2\ldots,K$}
	\State $\ten{G}(n)_{-k} \gets \ten{X}(n)\mmprs{\mat{U}}{k} \ \forall n$
	\State $\mat{S}_{-k,\text{w}} \gets
		\textstyle{\sum_n^N}\tilde{\mat{G}}_{-k,k}(n)\cdot\tilde{\mat{G}}_{-k,k}^\intercal(n)$
	\State $\mat{S}_{-k,\text{b}} \gets
		\textstyle{\sum_c^C}N_c\tilde{\bar{\mat{G}}}_{-k,k}(c)\cdot\tilde{\bar{\mat{G}}}_{-k,k}^\intercal(c)$
	\State $\varphi_k\gets\tr\left(\mat{U}_k^\intercal \mat{S}_{-k,\text{b}}\mat{U}_k\right)/\tr\left(\mat{U}_k^\intercal\mat{S}_{-k\text{w}}\mat{U}_k\right)$
	\State $\mat{V}_k\gets$ \parbox[t]{5cm}{largest magnitude $R_k$ \\ eigenvectors of
	$\mat{S}_{-k,\text{b}} - \varphi_k\mat{S}_{-k,\text{w}}$}
	\State $\mat{U}_k \gets$ \parbox[t]{5cm}{largest magnitude $R_k$ \\
	eigenvectors of $\mat{V}_k\mat{V}_k^\intercal\mat{S}_{k,\text{t}}\mat{V}_k\mat{V}_k^\intercal$}
	\EndFor
	\State $i\gets i+1$
	\Until{$i=I_\text{max}$ or $||\mat{U}_k^{(i)}-\mat{U}_k^{(i-1)}||<\epsilon
		\ \forall k$}
\end{algorithmic}

%% file: hoda_fw.tikz.tex
\footnotesize
\begin{tikzpicture}[y=-1cm]
	\TensorThree{$\ten{X}$}{$D_1$}{$D_2$}{$D_3$}{2}{2}{2}
	\node at (3.25,.5){$=$};

	\begin{scope}[shift={(6.25,0)}]
		\begin{scope}[shift={(-0.0707, -0.0707)}]
			\MatrixSkewed{$\mat{A}_1$}{$R_1$}{$D_1$}{2}{1}
		\end{scope}
		\begin{scope}[shift={(-0.1,0)}]
			\MatrixLeft{$\mat{A}_2$}{$R_2$}{$D_2$}{2}{1}
		\end{scope}
		\begin{scope}[shift={(0,1.1)}]
			\MatrixBelow{$\mat{A}_3$}{$R_3$}{$D_3$}{1}{2}
		\end{scope}
		\TensorThree{$\ten{G}$}{}{}{}{1}{1}{1}

		\node at (1.75,.5){$+$};

		\begin{scope}[shift={(2.25,0)}]
			\TensorThree{$\ten{E}$}{}{}{}{2}{2}{2}
		\end{scope}

	\end{scope}

\end{tikzpicture}

%% file: alg_hoda_fw.tex
\begin{algorithmic}[1]
	\Require $\{\ten{G}(n)\}_n^N,\{\ten{X}(n)\}_n^N,I_\text{max}, \epsilon$
	\State $\mat{A}_k \gets $ $\mat{U}_k \ \forall k$
	\State $i\gets 1$
	\Repeat
	\For{$k=1,2\ldots,K$}
	\State $\mat{X}_{-k}(n)\gets\ten{G}(n)\mmprsi{\mat{A}}{k} \ \forall n$
	\State
	$\mat{A}_k\gets\textstyle{\argmin_{\mat{A}_k}\sum_n^N}\left[\mat{X}_k(n)-\mat{A}_k\mat{X}_{-k}(n)\right]^2$
	\EndFor
	\State $i\gets i+1$
	\Until{$i=I_\text{max}$ or $||\mat{A}_k^{(i)}-\mat{A}_k^{(i-1)}||<\epsilon\
		\forall k$}
\end{algorithmic}

%% file: bttda_fw.tikz.tex
\footnotesize
\begin{tikzpicture}[y=-1cm]
	\useasboundingbox (0,-2) rectangle (16.5,3.5);
	\TensorThree{$\ten{X}$}{$D_1$}{$D_2$}{$D_3$}{2}{2}{2}

	\node at (3.15,0.5,0) {$=$};

	\begin{scope}[shift={(6.2,0)}]
		\begin{scope}[shift={(-0.0707, -0.0707)}]
			\MatrixSkewed{$\mat{A}_1^{(1)}$}{$R_1^{(1)}$}{$D_1$}{2}{1}
		\end{scope}
		\begin{scope}[shift={(-0.1,0)}]
			\MatrixLeft{$\mat{A}_2^{(1)}$}{$R_2^{(1)}$}{$D_2$}{2}{1}
		\end{scope}
		\begin{scope}[shift={(0,1.1)}]
			\MatrixBelow{$\mat{A}_3^{(1)}$}{$R_3^{(1)}$}{$D_3$}{1}{2}
		\end{scope}
		\TensorThree{$\ten{G}^{(1)}$}{}{}{}{1}{1}{1}
	\end{scope}

	\node at (8.30,0.5,0) {$+\cdots+$};

	\begin{scope}[shift={(11.85,0)}]
		\begin{scope}[shift={(-0.0707, -0.0707)}]
			\MatrixSkewed{$\mat{A}_1^{(B)}$}{$R_1^{(B)}$}{$D_1$}{2}{1}
		\end{scope}
		\begin{scope}[shift={(-0.1,0)}]
			\MatrixLeft{$\mat{A}_2^{(B)}$}{$R_2^{(B)}$}{$D_2$}{2}{1}
		\end{scope}
		\begin{scope}[shift={(0,1.1)}]
			\MatrixBelow{$\mat{A}_3^{(B)}$}{$R_3^{(B)}$}{$D_3$}{1}{2}
		\end{scope}
		\TensorThree{$\ten{G}^{(B)}$}{}{}{}{1}{1}{1}
	\end{scope}

	\node at (13.6,0.5,0) {$+$};

	\begin{scope}[shift={(14,0)}]
		\TensorThree{$\ten{E}^{(B)}$}{}{}{}{2}{2}{2}
	\end{scope}

\end{tikzpicture}

%% file: alg_bttda.tex
\begin{algorithmic}[1]
	\Require $\{\ten{X}(n)\}_n^N, \{c_n\}_n^N, \{(R_1^{(b)},R_2^{(b)},\ldots,R_K^{(b)})\}_b^B$
	\State $\ten{E}(n)\gets\ten{X}(n)\ \forall n$
	\For{$b=1,2,\ldots,B$}
	\State $\{\mat{U}^{(b)}\}\gets$ \parbox[t]{5cm}{\textsc{hoda} on $\{\ten{E}(n)\}_n^N$ and
	$\{c_n\}_n^N$ with rank $(R_1^{(b)},R_2^{(b)},\ldots,R_K^{(b)})$}
	\State $\ten{G}^{(b)}(n)\gets\ten{E}(n)\mmpri{\mat{U}^{(b)}}\
		\forall n$
	\State $\{\mat{A}^{(b)}\}\gets$ \parbox[t]{5cm}{Forward \textsc{hoda} on
		$\{\ten{G}^{(b)}(n)\}_n^N$ and $\ten{E}$}
	\State
	$\hat{\ten{E}}(n)\gets\ten{G}^{(b)}(n)\mmpri{\mat{A}^{\intercal(b)}}\
		\forall n$
	\State
	$\ten{E}(n)\gets \ten{E}(n) - \hat{\ten{E}}(n) \forall n$

	\EndFor
\end{algorithmic}

%% file: pairwise_erp.tikz.tex
\scriptsize\noindent\begin{tikzpicture}
	\pgfplotstableread[col sep=comma]{erp_BTTDA_HODA.csv}\datatable
	\begin{groupplot}[
			group style={%
					group size=3 by 1,
					horizontal sep=1em
				},
			error bars/x dir=both,
			error bars/x explicit,
			error bars/error mark=none,
			xmin=-0.25, xmax=2.75,
			yticklabels from table={\datatable}{dataset},
			ytick=data,
			enlarge y limits=0.05,
			width=\textwidth/3-11em/3,
			height=0.25\textwidth,
			scale only axis,
			point meta=explicit symbolic,
			nodes near coords,
			every node near coord/.append style={
					anchor=west,
					yshift=2,
					font=\scriptsize,
				},
		]

		\newcommand{\addPlotComparison}[2]{
			\nextgroupplot[title={$\leftarrow$ #2 / #1 $\rightarrow$}]
			\pgfplotstableread[col sep=comma]{erp_#1_#2.csv}\datatable
			\addplot+ table[
					only marks,
					x=smd,
					y=dataset_idx,
					meta=p_star,
					x error=ci,
					nodes near coords style={
							font=\tiny,
						}
				] {\datatable};
			\draw (axis cs:0, \pgfkeysvalueof{/pgfplots/ymin}) -- (axis cs:0, \pgfkeysvalueof{/pgfplots/ymax});
		}

		\addPlotComparison{BTTDA}{HODA}
		\addPlotComparison{PARAFACDA}{HODA}
		\addPlotComparison{BTTDA}{PARAFACDA}

	\end{groupplot}
\end{tikzpicture}

%% file: pairwise_mi.tikz.tex
\scriptsize\noindent\begin{tikzpicture}
	\pgfplotstableread[col sep=comma]{mi_BTTDA_HODA.csv}\datatable
	\begin{groupplot}[
			group style={%
					group size=3 by 1,
					horizontal sep=1em
				},
			xlabel={standardized mean difference},
			error bars/x dir=both,
			error bars/x explicit,
			error bars/error mark=none,
			xmin=-5.5, xmax=6.5,
			yticklabels from table={\datatable}{dataset},
			ytick=data,
			enlarge y limits=0.15,
			width=\textwidth/3-11em/3,
			height=6/15*0.25\textwidth,
			scale only axis,
			point meta=explicit symbolic,
			nodes near coords,
			every node near coord/.append style={
					anchor=west,
					yshift=2,
					font=\scriptsize,
				},
		]

		\newcommand{\addPlotComparison}[2]{
			\nextgroupplot
			\pgfplotstableread[col sep=comma]{mi_#1_#2.csv}\datatable
			\addplot+ table[
					only marks,
					x=smd,
					y=dataset_idx,
					meta=p_star,
					x error=ci,
					nodes near coords style={
							font=\tiny,
						}
				] {\datatable};
			\draw (axis cs:0, \pgfkeysvalueof{/pgfplots/ymin}) -- (axis cs:0, \pgfkeysvalueof{/pgfplots/ymax});
		}

		\addPlotComparison{BTTDA}{HODA}
		\addPlotComparison{PARAFACDA}{HODA}
		\addPlotComparison{BTTDA}{PARAFACDA}

	\end{groupplot}
\end{tikzpicture}

%% file: score_erp.tex
\begin{tabular}{@{}lrrrrrrrrrrrrrrr@{}}
\toprule
Pipelines & BNCI2014-008 & BNCI2014-009 & BNCI2015-003 & BrainInvaders2012 & BrainInvaders2013a \\
\midrule
ERPCov+MDM & $74.30\pm9.77$ & $81.16\pm10.13$ & $76.79\pm10.95$ & $78.77\pm10.32$ & $80.59\pm9.36$ \\
ERPCov(svdn4)+MDM & $75.42\pm9.91$ & $84.52\pm8.83$ & $76.93\pm11.26$ & $79.02\pm10.53$ & $82.07\pm8.46$ \\
XDAWN+LDA & $82.24\pm5.26$ & $64.03\pm3.91$ & $78.62\pm7.19$ & $64.41\pm4.14$ & $76.74\pm7.16$ \\
XDAWNCov+MDM & $77.62\pm9.81$ & $92.04\pm5.97$ & $83.08\pm7.55$ & $88.22\pm5.90$ & $90.97\pm5.52$ \\
XDAWNCov+TS+SVM & $85.61\pm4.43$ & $93.43\pm5.11$ & $82.95\pm8.57$ & $90.99\pm4.79$ & \boldmath$92.71\pm4.92$ \\HODA+LDA & $85.20\pm4.62$ & $93.23\pm4.07$ & $82.74\pm7.14$ & $86.74\pm5.33$ & $89.78\pm6.12$ \\
PARAFACDA+LDA & $85.99\pm4.73$ & $94.03\pm4.92$ & $84.71\pm7.54$ & $90.80\pm4.87$ & $91.78\pm5.81$ \\
BTTDA+LDA & \boldmath$86.20\pm4.61$ & \boldmath$94.40\pm4.59$ & \boldmath$84.99\pm7.37$ & \boldmath$91.27\pm4.40$ & $92.08\pm5.50$ \\
\midrule 
Pipelines & BrainInvaders2014a & BrainInvaders2014b & BrainInvaders2015a & BrainInvaders2015b & Cattan2019-VR \\
\midrule
ERPCov+MDM & $71.62\pm11.17$ & $78.57\pm12.36$ & $80.02\pm10.07$ & $75.04\pm15.85$ & $80.76\pm10.07$ \\
ERPCov(svdn4)+MDM & $72.11\pm11.64$ & $76.48\pm12.83$ & $77.92\pm10.33$ & $77.09\pm15.81$ & $80.67\pm9.47$ \\
XDAWN+LDA & $66.60\pm7.54$ & $83.73\pm10.62$ & $76.02\pm10.46$ & $77.22\pm13.73$ & $67.16\pm6.11$ \\
XDAWNCov+MDM & $80.88\pm11.01$ & $91.58\pm10.02$ & $92.57\pm5.03$ & $83.48\pm12.05$ & $88.53\pm7.34$ \\
XDAWNCov+TS+SVM & $85.77\pm9.75$ & \boldmath$91.88\pm9.94$ & $93.05\pm4.98$ & \boldmath$84.56\pm12.09$ & $90.68\pm6.29$ \\HODA+LDA & $83.76\pm9.50$ & $87.23\pm11.00$ & $90.96\pm5.68$ & $82.19\pm11.68$ & $88.21\pm8.59$ \\
PARAFACDA+LDA & $87.24\pm9.55$ & $90.00\pm9.70$ & $92.95\pm4.51$ & $84.33\pm12.39$ & $91.22\pm7.95$ \\
BTTDA+LDA & \boldmath$87.71\pm9.36$ & $90.41\pm10.59$ & \boldmath$93.44\pm4.38$ & $84.47\pm12.26$ & \boldmath$91.47\pm7.37$ \\

\midrule 
Pipelines & EPFLP300 & Huebner2017 & Huebner2018 & Lee2019-ERP & Average \\
\midrule
ERPCov+MDM & $71.97\pm10.88$ & $94.47\pm8.26$ & $95.15\pm3.72$ & $74.43\pm13.26$ & $79.55\pm10.76$ \\
ERPCov(svdn4)+MDM & $71.44\pm10.20$ & $96.21\pm6.50$ & $96.61\pm1.89$ & $82.47\pm12.56$ & $80.64\pm10.49$ \\
XDAWN+LDA & $62.98\pm5.38$ & $97.74\pm2.84$ & $97.54\pm1.58$ & $96.45\pm3.93$ & $77.96\pm7.19$ \\
XDAWNCov+MDM & $83.20\pm9.05$ & $98.07\pm2.09$ & $97.78\pm1.04$ & $97.70\pm2.68$ & $88.98\pm7.53$ \\
XDAWNCov+TS+SVM & $84.29\pm8.53$ & \boldmath$98.69\pm1.78$ & \boldmath$98.47\pm0.97$ & \boldmath$98.41\pm2.03$ & $90.82\pm6.82$ \\HODA+LDA & $82.09\pm8.11$ & $97.70\pm3.09$ & $97.61\pm1.57$ & $97.00\pm2.72$ & $88.89\pm7.04$ \\
PARAFACDA+LDA & $85.13\pm8.72$ & $98.52\pm1.86$ & $98.36\pm0.93$ & $98.09\pm1.84$ & $90.94\pm6.90$ \\
BTTDA+LDA & \boldmath$85.95\pm8.06$ & $98.54\pm1.84$ & $98.40\pm0.90$ & $98.13\pm1.83$ & \boldmath$91.25\pm6.77$ \\
\bottomrule
\end{tabular}

%% file: score_mi.tex
\begin{tabular}{@{}lrrrrrr@{}}
\toprule
Pipelines & AlexandreMotorImagery & BNCI2014-001 & Schirrmeister2017 & Weibo2014 & Zhou2016 \\
\midrule
ACM+TS+SVM & $69.37\pm15.07$ & \boldmath$77.82\pm12.23$ & $82.50\pm10.20$ & \boldmath$63.89\pm11.01$ & \boldmath$85.25\pm4.06$ \\
EEGTCNet & $34.17\pm1.86$ & $41.65\pm13.73$ & $71.11\pm11.96$ & $17.95\pm3.88$ & $37.19\pm2.57$ \\
FgMDM & $65.63\pm15.63$ & $70.14\pm15.13$ & $82.97\pm10.08$ & $56.94\pm9.26$ & $83.07\pm4.96$ \\
ShallowConvNet & $50.00\pm12.94$ & $72.47\pm16.50$ & $85.13\pm9.57$ & $48.94\pm10.36$ & $85.02\pm3.78$ \\
TS+EL & \boldmath$69.79\pm13.75$ & $72.38\pm14.85$ & \boldmath$85.53\pm9.40$ & $63.84\pm8.77$ & $84.54\pm4.93$ \\HODA+LDA & $50.00\pm15.17$ & $53.84\pm12.46$ & $72.18\pm9.02$ & $54.69\pm10.53$ & $74.27\pm6.27$ \\
PARAFACDA+LDA & $46.67\pm12.22$ & $53.49\pm12.28$ & $76.11\pm13.29$ & $53.49\pm11.04$ & $64.68\pm6.00$ \\
BTTDA+LDA & $49.58\pm15.68$ & $57.42\pm13.96$ & $79.24\pm12.51$ & $59.36\pm11.61$ & $77.02\pm4.04$ \\
\midrule 
Pipelines & Average \\
\midrule
ACM+TS+SVM & \boldmath$75.77\pm11.12$ \\
EEGTCNet & $40.41\pm8.45$ \\
FgMDM & $71.75\pm11.71$ \\
ShallowConvNet & $68.31\pm11.43$ \\
TS+EL & $75.22\pm10.95$ \\HODA+LDA & $61.00\pm11.11$ \\
PARAFACDA+LDA & $58.89\pm11.27$ \\
BTTDA+LDA & $64.52\pm12.23$ \\
\bottomrule
\end{tabular}

%% file: gridsearch.tikz.tex
\scriptsize
\noindent\begin{tikzpicture}[trim axis group left]
	\begin{groupplot}[
			group style={%
					group size=2 by 1,
					x descriptions at=edge bottom,
					horizontal sep=4.5em
				},
			width=\linewidth/2-4.5em/2,
			scale only axis,
			xlabel=$b$,
		]

		\newcommand{\addPlotTheta}[2]{%
			\addplot+ table [
					color=red,
					x=n_blocks,
					y=#1,
					col sep=comma,
					restrict expr to domain={\thisrow{theta}}{#2:#2},
					unbounded coords=discard, 
				]{gridsearch_erp_processed.csv};
		}

		\nextgroupplot[
			legend style={at={(0.98,0.02)},anchor=south east},
			ylabel=ROC-AUC,
		]
		\addPlotTheta{test_score}{0}
		\addlegendentry{$\theta=0.0$}
		\addPlotTheta{test_score}{0.1}
		\addlegendentry{$\theta=0.1$}
		\addPlotTheta{test_score}{1}
		\addlegendentry{$\theta=1.0$}
		\addplot[mark=none,brown!60!black] coordinates {(1,0.8540042316897378) (16,0.8540042316897378)};

		\nextgroupplot[
			ymin=0.9475, ymax=1,
			ylabel=NMSE,
			ytick={0.95, 0.97, 0.99},
			yticklabels={0.95, 0.97, 0.99}
		]
		\addPlotTheta{test_mse}{0}
		\addPlotTheta{test_mse}{0.1}
		\addPlotTheta{test_mse}{1}

	\end{groupplot}
\end{tikzpicture}

%% file: erp_contrast.tikz.tex
\scriptsize
\begin{tikzpicture}[trim axis group left]

	\begin{groupplot}[
			group style={
					group size=6 by 1,
					horizontal sep=0.2cm,
				},
			width=\linewidth/4,
			ylabel={Amplitude ($\mu V$)},
			legend pos=north east,
			no markers,
			enlarge x limits=false,
			xmin=-0.2,
			xmax=1.0,
			ymin=-2.5,
			ymax=2.5
		]

		\newcommand{\drawEvokedTime}[1]{
			\pgfplotstableread[col sep=comma]{erp_contrast_#1.csv}\datatable
			\pgfplotstablegetcolsof{\datatable}
			\pgfmathtruncatemacro{\numcols}{\pgfplotsretval-1}
			\pgfplotsinvokeforeach {1,...,\numcols} {
				\addplot+[
					color=blue!80!black,
					no markers,
					thin,
					solid
				] table [
						x=time,
						y index=##1,
						col sep=comma,
					] {erp_contrast_#1.csv};
			}
			\addplot+[color=black, dashed] coordinates {(-0.2,0) (1.0,0)};
			\addplot+[color=black, dashed] coordinates {(0, -2.5) (0,2.5)};
		}

		\nextgroupplot[title=Grand-average contrast]
		\addplot+[color=red] coordinates {(0.45, -2.5) (0.45, 2.5)};
		\drawEvokedTime{grand-avg}
		\nextgroupplot[group/empty plot]
		\addplot graphics[xmin=-0.1, xmax=0.9, ymin=-2.5, ymax=2.5]{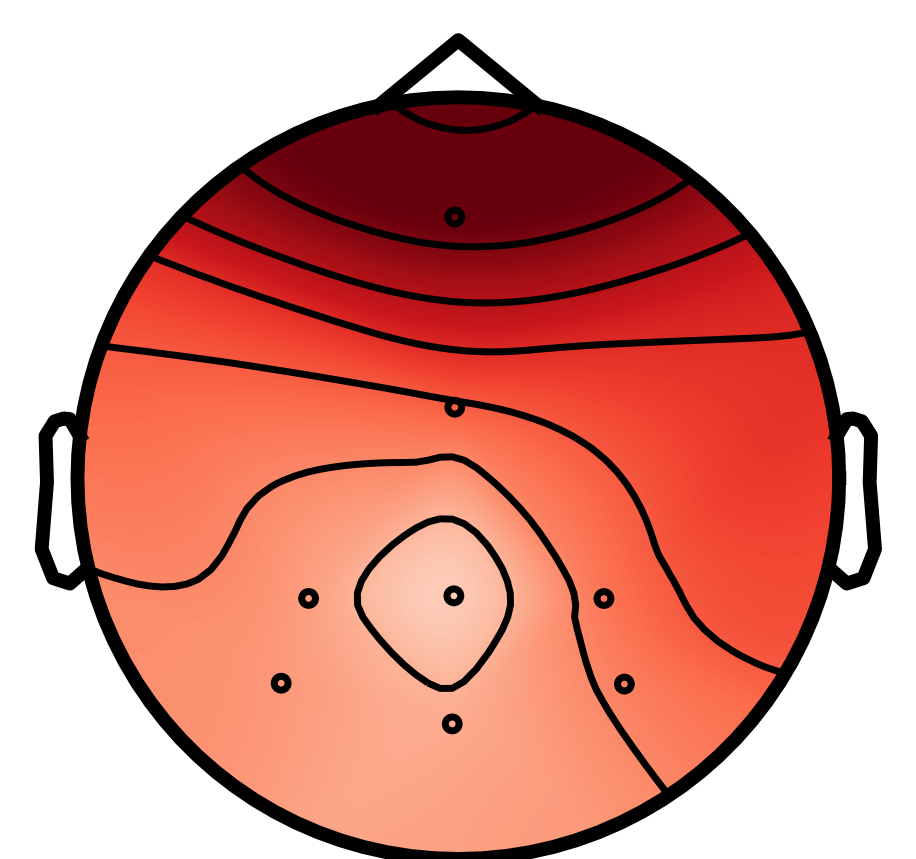};

		\nextgroupplot[title={Reconstructed contrast ($b=1$)}, extra x ticks={0.3}]

		\addplot+[color=red] coordinates {(0.30, -2.5) (0.30, 2.5)};

		\node[coordinate, pin={%
		[pin distance=1mm, inner sep=0.5pt, text depth=0pt] above:{\tiny P1}
		}] at (axis cs:0.135,0.95) {};
		\node[coordinate, pin={%
		[pin distance=1mm, inner sep=0.5pt, text depth=0pt] below left:{\tiny N1}
		}] at (axis cs:0.2175,-1.32) {};
		\node[coordinate, pin={%
		[pin distance=1mm, inner sep=0.5pt, text depth=0pt] right:{\tiny N2}
		}] at (axis cs:0.3,-1.575) {};

		\drawEvokedTime{block-1}

		\nextgroupplot[group/empty plot]
		\addplot graphics[xmin=-0.1, xmax=0.9, ymin=-2.5, ymax=2.5]{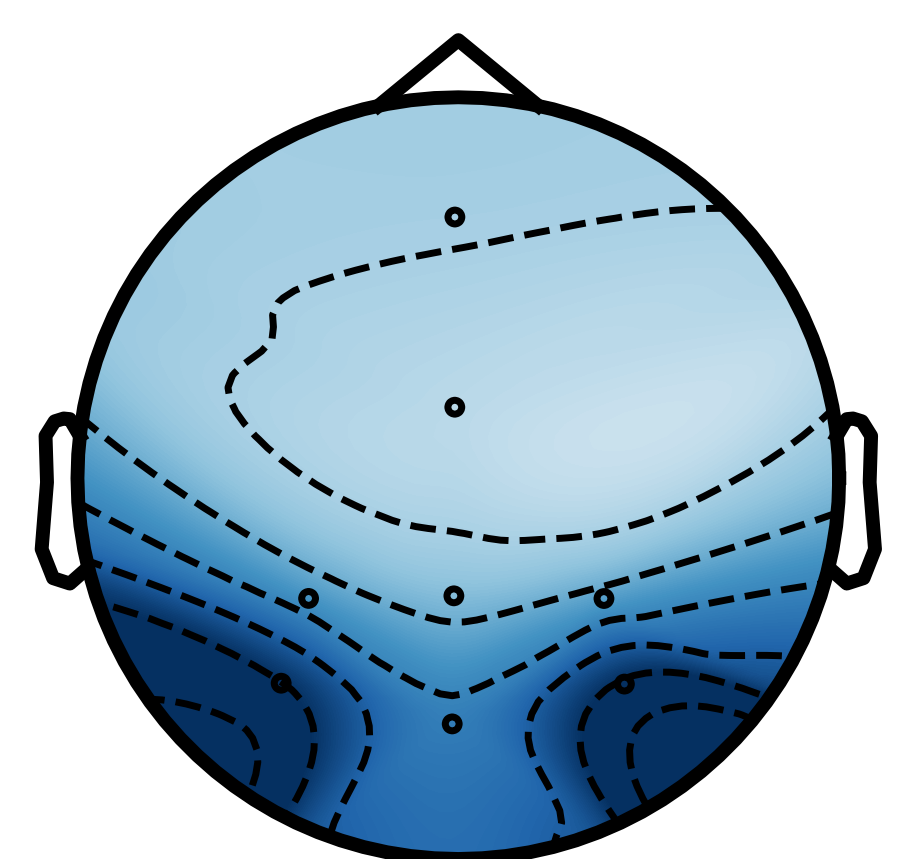};

		\nextgroupplot[title={Reconstructed contrast ($b=2$)}, extra x ticks={0.3}]
		\addplot+[color=red] coordinates {(0.30, -2.5) (0.30, 2.5)};

		\node[coordinate, pin={%
		[pin distance=1mm, inner sep=0.5pt, text depth=0pt] above left:{\tiny P3a}
		}] at (axis cs:0.2,0.75) {};
		\node[coordinate, pin={%
		[pin distance=1mm, inner sep=0.5pt, text depth=0pt] above right:{\tiny P3b}
		}] at (axis cs:0.3,1.15) {};

		\drawEvokedTime{block-2}
		\nextgroupplot[group/empty plot]
		\addplot graphics[xmin=-0.1, xmax=0.9, ymin=-2.5, ymax=2.5]{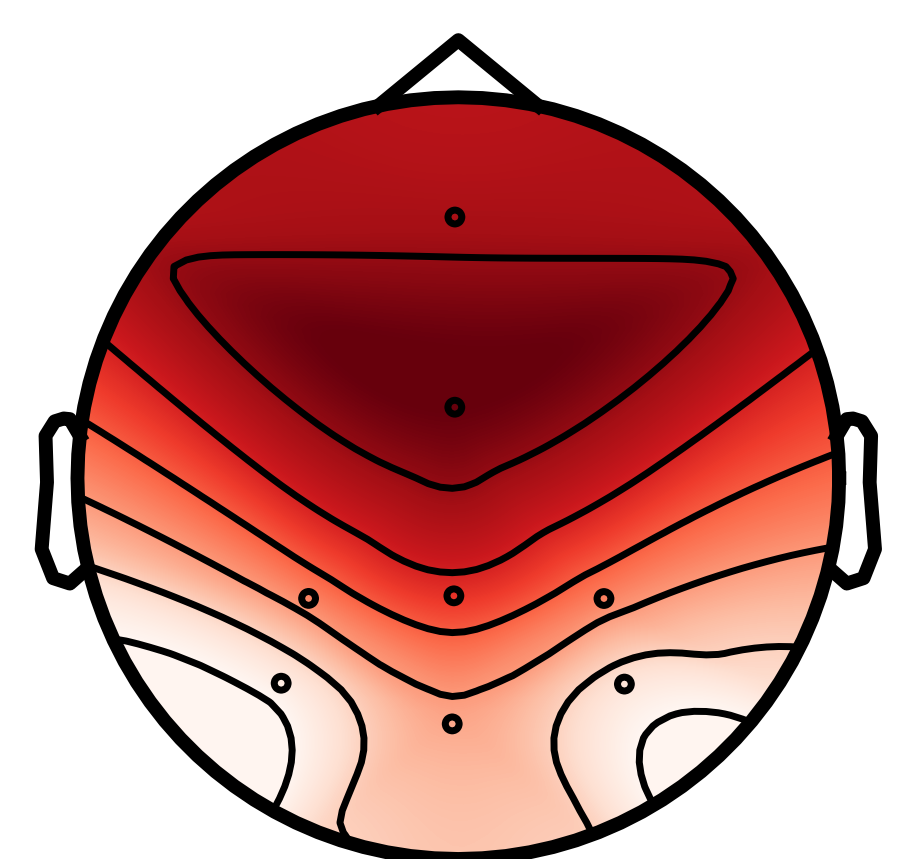};

	\end{groupplot}
\end{tikzpicture}

%% file: mi_contrast.tikz.tex
\scriptsize
\begin{tikzpicture}[trim axis group left]

	\begin{groupplot}[
			group style={
					group size=6 by 1,
					horizontal sep=0.2cm,
				},
			width=\linewidth/4,
			xlabel={Time (s)},
			ylabel={Frequency (Hz)},
			ymin=8,
			ymax=32,
			xmin=0,
			xmax=3,
			ymode=log,
			ytick={8,16,32},
			log ticks with fixed point,
			axis on top,
			no markers,
		]

		\nextgroupplot[extra y ticks={11}]
		\addplot graphics[xmin=0, xmax=3, ymin=8, ymax=32]{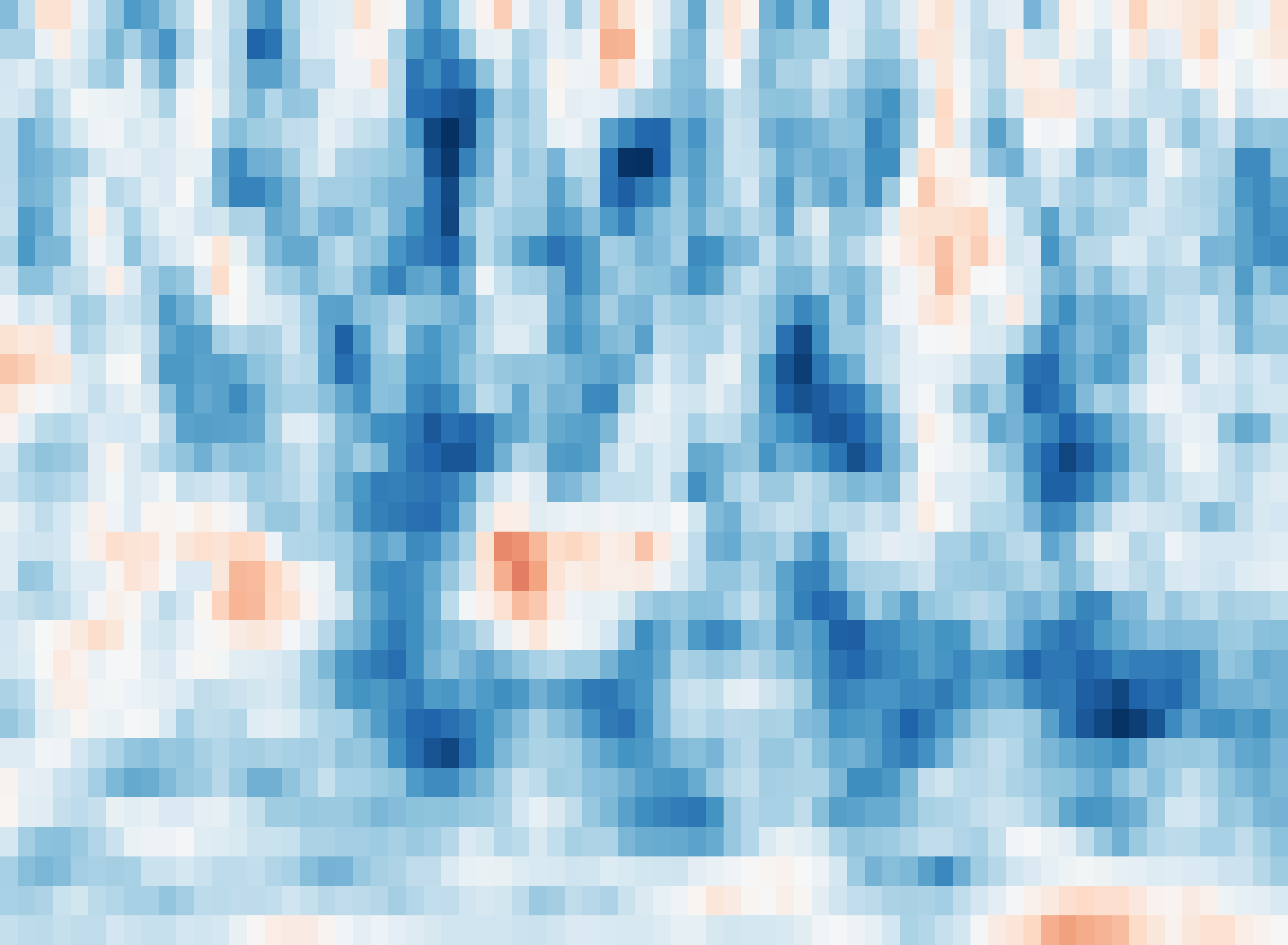};
		\addplot+[color=red] coordinates {(0, 11) (3, 11)};
		\addplot+[color=red] coordinates {(1, 8) (1, 32)};

		\nextgroupplot[group/empty plot]
		\addplot graphics[xmin=0.3, xmax=2.7, ymin=8, ymax=32]{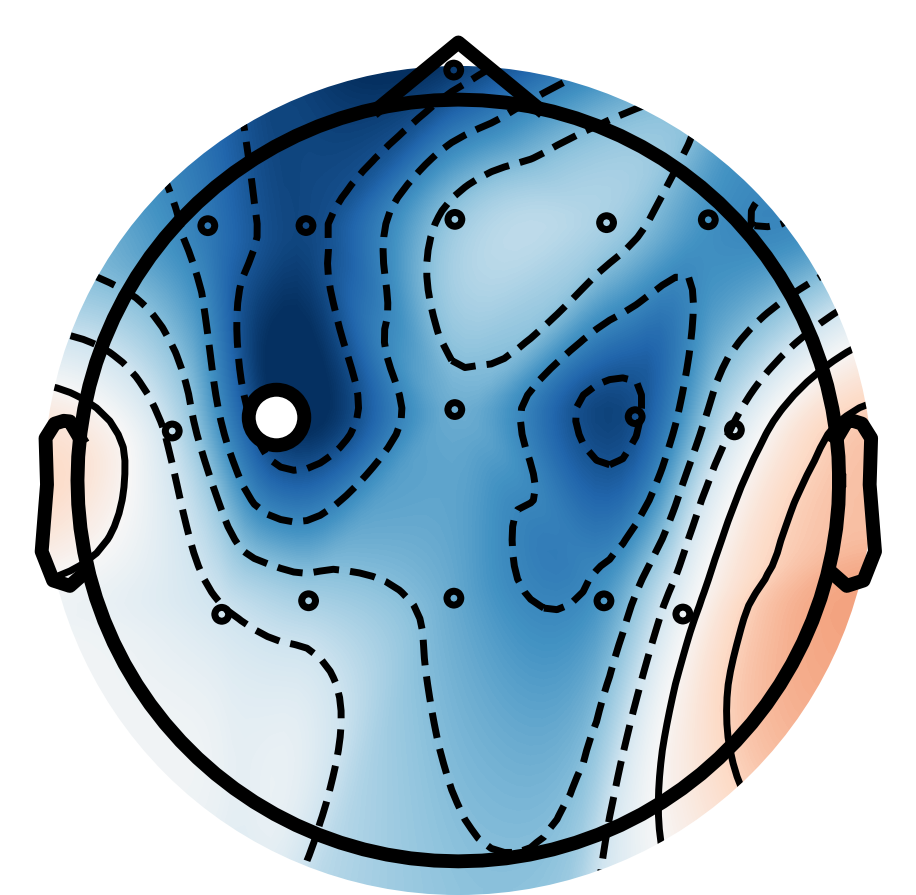};

		\nextgroupplot[extra y ticks={11}]
		\addplot graphics[xmin=0, xmax=3, ymin=8, ymax=32]{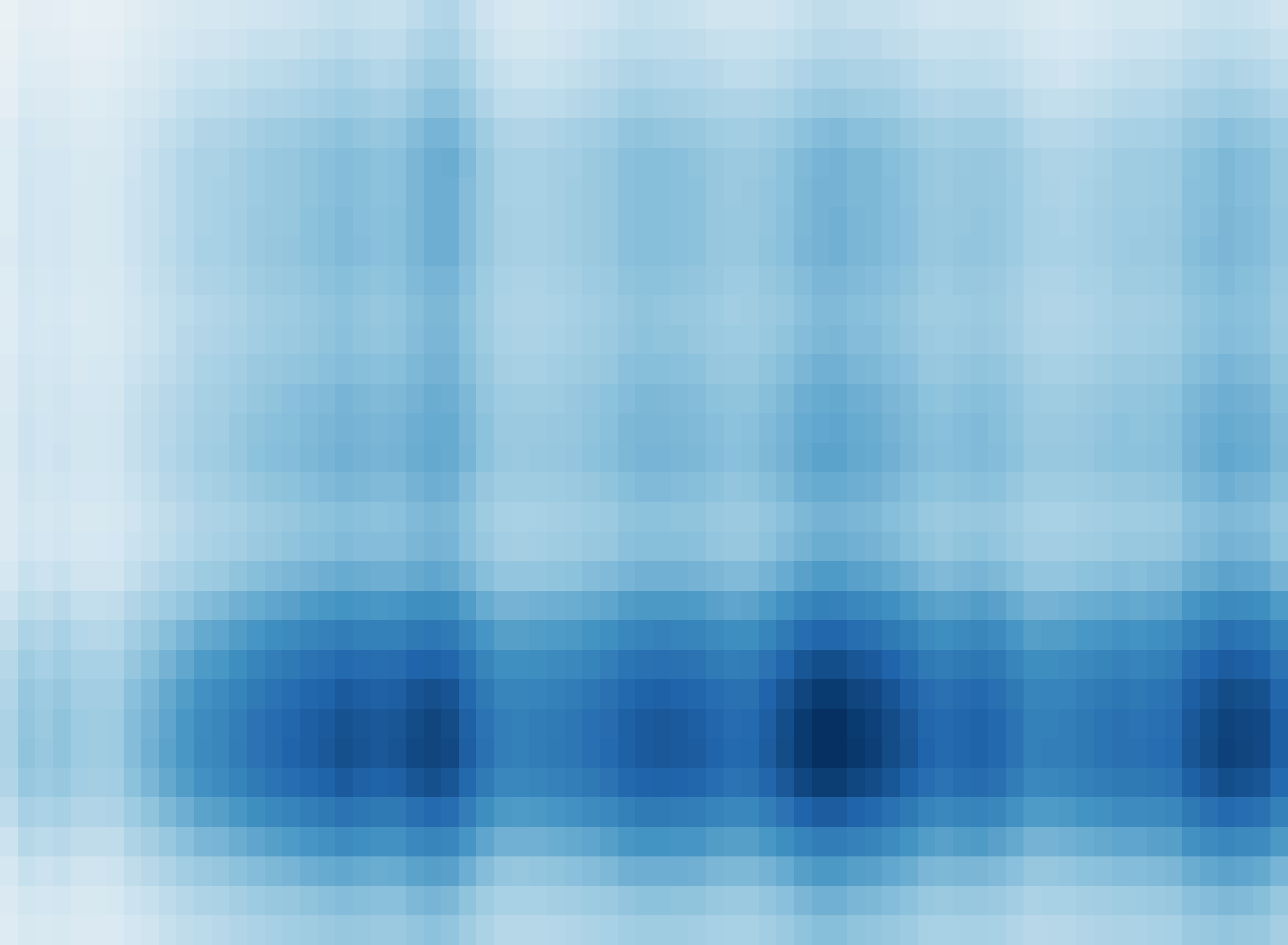};
		\addplot+[color=red] coordinates {(0, 11) (3, 11)};
		\addplot+[color=red] coordinates {(1, 8) (1, 32)};
		\nextgroupplot[group/empty plot]
		\addplot graphics[xmin=0.3, xmax=2.7, ymin=8, ymax=32]{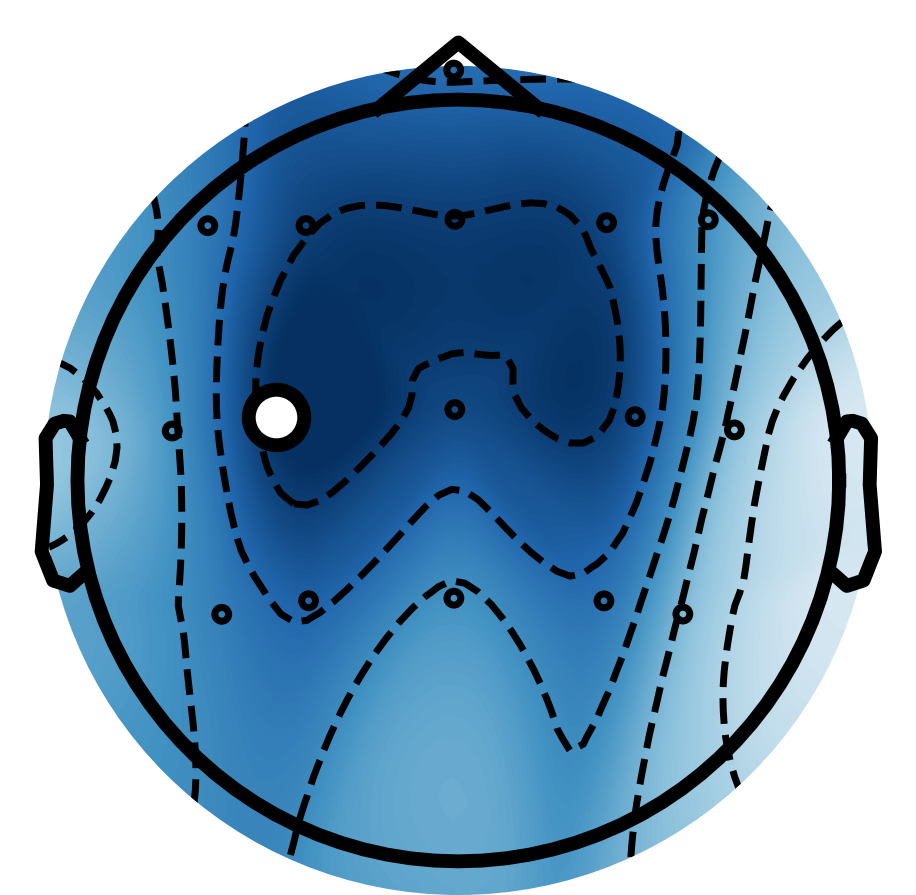};

		\nextgroupplot[extra x ticks = {1.6}, extra y ticks={10}]
		\addplot graphics[xmin=0, xmax=3, ymin=8, ymax=32]{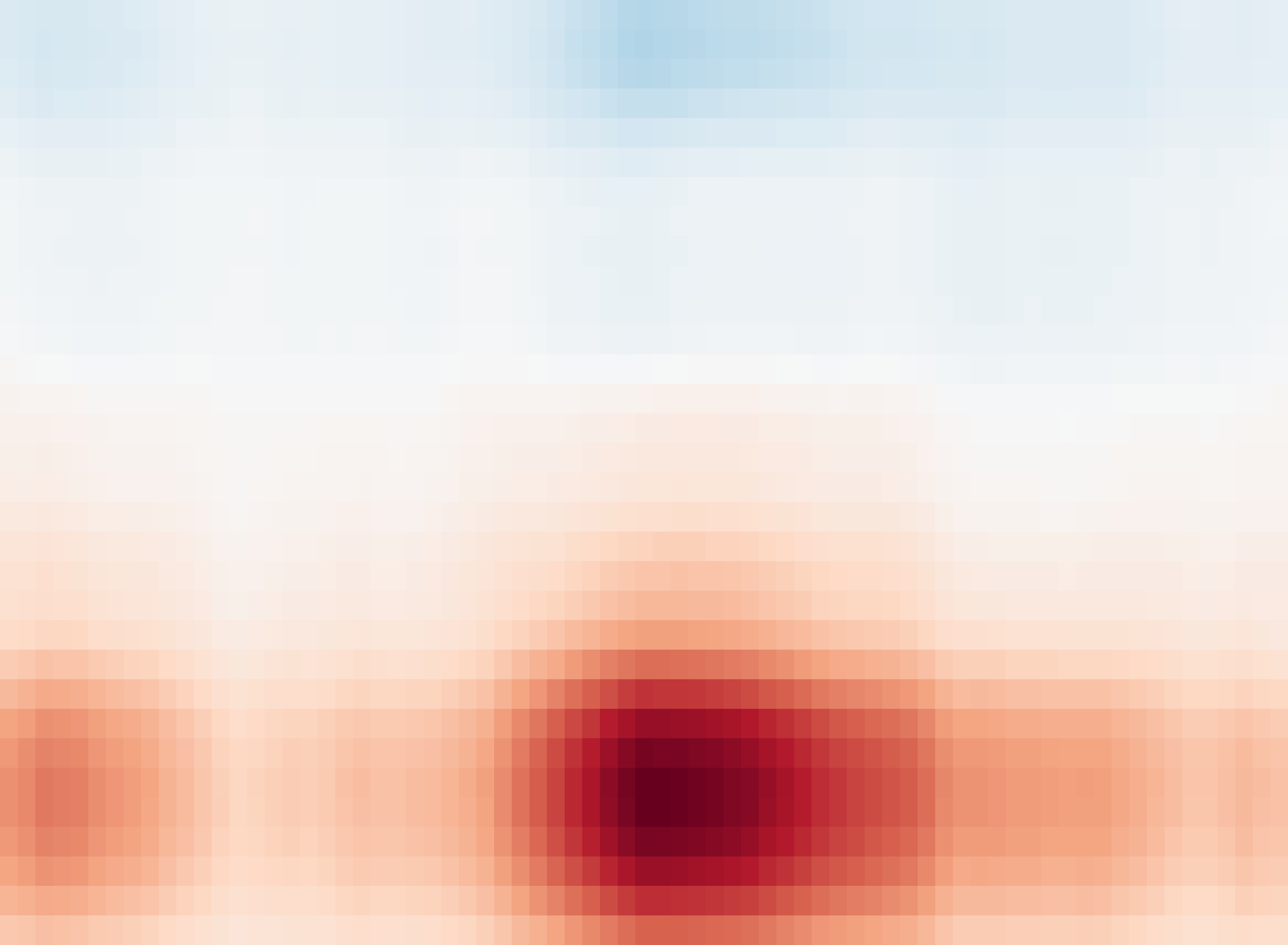};
		\addplot+[color=red] coordinates {(0, 10) (3, 10)};
		\addplot+[color=red] coordinates {(1.6, 8) (1.6, 32)};

		\nextgroupplot[group/empty plot]
		\addplot graphics[xmin=0.3, xmax=2.7, ymin=8, ymax=32]{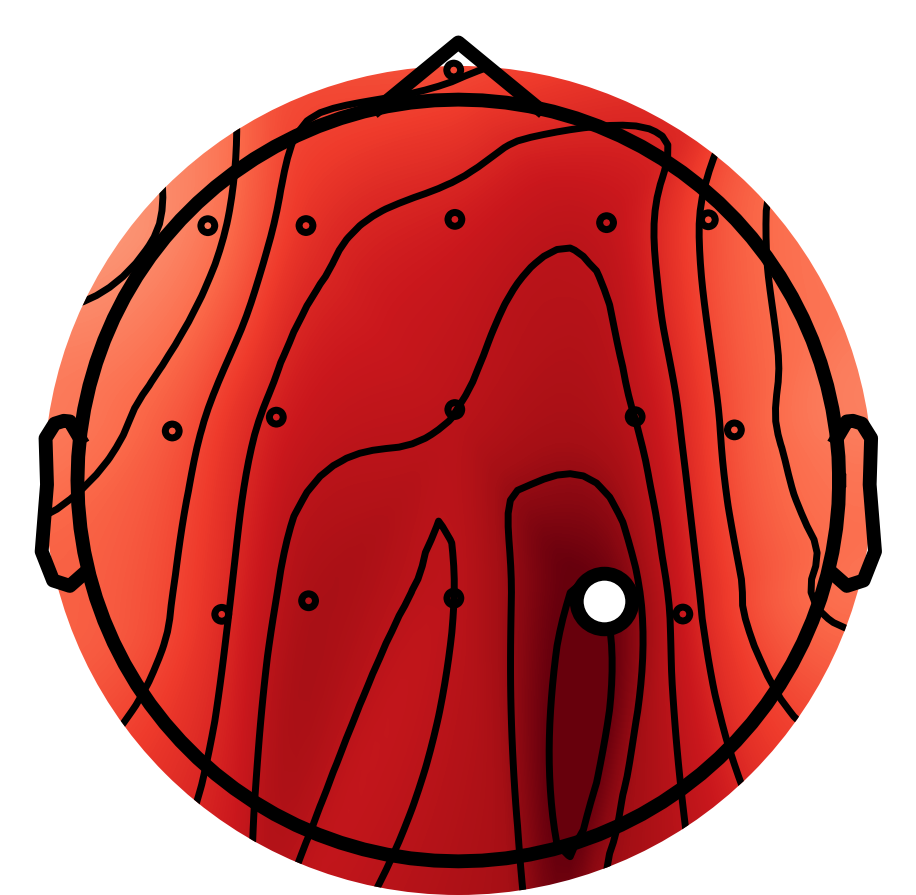};

	\end{groupplot}
\end{tikzpicture}

%% file: complexity_derivation.tex
\section{Time complexity of HODA and BTTDA}
\label{app:complexity-derivation}

\subsection{Backward \Ac{hoda} algorithm}
To obtain time complexity of the \ac{hoda} backward model algorithm, we start
by determining the number of operations within a single inner loop
for given $i$ and $k$.
Since we are interested in worst-case complexity, assume all input dimensions $D_k$
are equal to $\max_k(D_k) = D$ and all reduced dimensions $R_k$ take their maximum
value $R_k=D_k = D$.

The multi-mode products $\smash{\ten{X}(n)\mmprsi{U}{k}}$ require
\begin{equation}
	\begin{split}
		\left(K-1\right)\left[DD\left(ND^{K-1}\right)\right] = N\left(K-1\right)D^{K+1}
	\end{split}
\end{equation}
operations.
Calculating class means and centering the data requires
\begin{equation}
	\begin{split}
		& C\left(DD^{K-1} + D + DD^{K-1}\right) \\
		& = ND^K + D + ND^K \\
		& = 2ND^K + D
	\end{split}
\end{equation}
From this, the number of operations for the within-subject scatter matrix $\mat{S}_{-k, \text{w}}$
\begin{equation}
	\begin{split}
		D(ND^{K-1})D = ND^{K+1}
	\end{split}
\end{equation}
and the between-class scatter matrix $\mat{S}_{-k, \text{b}}$
\begin{equation}
	\begin{split}
		C\left(D1D\right) + CD^2 = 2CD^2
	\end{split}
\end{equation}
can be obtained.
$\varphi_k$ can be calculated as $\tr\left(\mat{U}_k^\intercal \mat{S}_{-k,\text{b}}\mat{U}_k\right)/\tr\left(\mat{U}_k^\intercal\mat{S}_{-k\text{w}}\mat{U}_k\right)$ in
\begin{equation}
	\begin{split}
		& \left(DDD + DDD + D\right) + \left(DDD + DDD + D\right)+1 \\
		& =  2\left(2D^3 D\right) + 1
		= 4D^3+2D + 1
	\end{split}
\end{equation}
operations.
The difference of the scatter matrices
$\mat{S}_{-k, \text{w}} - \varphi - \mat{S}_{-k,\text{b}}$
then yields
\begin{equation}
	D^2 + D^2 + D^2 = 3D^2
\end{equation}
operations,
and its eigendecomposition
\begin{equation}
	D^3
\end{equation}
Finally,  the projection for orthogonalization
$\mat{V}_k\mat{V}_k^\intercal\mat{S}_{k,\text{t}}\mat{V}_k\mat{V}_k^\intercal$
adds
\begin{equation}
	DDD + DDD + DDD +DDD = 4D^3
\end{equation}
and its eigendecomposition
\begin{equation}
	D^3
\end{equation}

Together, this forms
\begin{equation}
	\begin{split}
		& N(K-1)D^{K+1} + 2ND^K + D + 2CD^2 + 4D^3 \\
		& \quad + 2D +1 +3D^2 + D^3 + 4D^3 + D^3 \\
		& = NKD^{K+1} -ND^{K+1} + 2ND^K + 10D^3 \\
		& \quad +(3+2C)D^2 + 3D+1
	\end{split}
\end{equation}
From the number of operations, the time complexity can be derived as
\begin{equation}
	\begin{split}
		& \mathcal{O}\left[ NKD^{K+1} -ND^{K+1} + 2ND^K + 10D^3\right. \\
		& \left. \quad +(3+2C)D^2 + 3D+1 \right] \\
		& = \mathcal{O}\left(NKD^{K+1} \right)
	\end{split}
\end{equation}
assuming $C < N$.
The procedure in the inner loop over $k$ and the outer loop over $i$ is
executed $I_\text{max}K$ times, yielding
\begin{equation}
	\begin{split}
		\mathcal{O}\left(I_\text{max}KNKD^{K+1}\right)
		= \mathcal{O}\left(I_\text{max}K^2ND^{K+1}\right)
	\end{split}
	\label{eq:complexity/backward}
\end{equation}

\subsection{Forward \ac{hoda} algorithm}
Similar to the previous derivation, we start by determining the operations
within a single iteration of a nested over the $i$ and $k$.

The first step is again a multi-mode product, $\smash{\ten{G}(n)\mmprsi{\mat{A}}{k}}$:
\begin{equation}
	\begin{split}
		\left(K-1\right)\left[DD\left(ND^{K-1}\right)\right] = N\left(K-1\right)D^{K+1}
	\end{split}
\end{equation}
The second step requires least squares regression which can be solved in
\begin{equation}
	\begin{split}
		& D\left(ND^{K-1}\right)D + D\left(ND^{K-1}\right)D + D^3 \\
		& = 2ND^{K+1} +D^3
	\end{split}
\end{equation}
operations.

Together, this forms
\begin{equation}
	\begin{split}
		& N(K-1)D^{K+1} + 2ND^{K+1} +D^3 \\
		& = NKD^{K+1} - ND^{K+1} + 2ND^{K+1} + D^3
	\end{split}
\end{equation}
The time complexity to fit one iteration of the algorithm for the forward model
is then
\begin{equation}
	\begin{split}
		& \mathcal{O}\left(NKD^{K+1} - ND^{K+1} + 2ND^{K+1} + D^3\right) \\
		& = \mathcal{O}\left(NKD^{K+1}\right)
	\end{split}
\end{equation}
and, when integrating it in the inner and outer loops over $i$ and $k$,
\begin{equation}
	\mathcal{O}\left(I_\text{max}KNKD^{K+1}\right)
	= \mathcal{O}\left(I_\text{max}NK^2D^{K+1}\right)
	\label{eq:complexity/forward}
\end{equation}

This is the same asymptotic time complexity as the backwards modeling algorithm,
since they are both dominated by the multi-mode product.

\subsection{Backward \ac{bttda} algorithm}

Fitting the \ac{bttda} model involves a loop over blocks $b$.
At each iteration, a backward model is fit with complexity as in
\cref{eq:complexity/backward}
The core tensors $\smash{\ten{G}^\text{(b)}(n)}$
are extracted with the multi-mode product using
\begin{equation}
	K\left[DD\left(ND^{K-1}\right)\right] = KND^{K+1}
	\label{eq:complexity/mmpr}
\end{equation}
operations

Next, the forward model is fit on these core tensors, with complexity as in
\cref{eq:complexity/forward}.
The number of steps for the reconstructed tensors can similarly be obtained using
\cref{eq:complexity/mmpr},
and calculating the residual requires
\begin{equation}
	ND^K
\end{equation}
operations.

A single block $b$ can thus be fit with complexity
\begin{equation}
	\begin{split}
		& \mathcal{O}\left(
		I_\text{max}NK^2D^{K+1}
		+ KND^{K+1} \right. \\
		& \left. \quad + I_\text{max}NK^2D^{K+1}
		+ KND^{K+1}
		+ ND^K
		\right) \\
		& = \mathcal{O}\left(I_\text{max}NK^2D^{K+1}\right)
	\end{split}
\end{equation}
The complexity when calculating all blocks is
\begin{equation}
	\mathcal{O}\left(BI_\text{max}NK^2D^{K+1}\right)
\end{equation}

\subsection{Hyperparameter tuning}
Finally, proper decoding training relies heavily on tuning the hyperparameters
$\theta$ and $B$ through cross-validation.
Let $F$ be the number of cross-validation folds and $\Theta$ a set of $\theta$
candidates.
We can take advantage of the fact that the \ac{bttda} model can be fit on a fixed
amount of blocks $B$, but intermediary blocks $1, 2, \cdots B$ can easily be extracted.
This way, no second iteration over candidates for $B$ is necessary and complexity
can be kept at
\begin{equation}
	\mathcal{O}\left(\left|\Theta\right|FBI_\text{max}NK^2D^{K+1}\right)
\end{equation}

%% file: moabb_datasets.tex
\begin{tabularx}{\linewidth}{@{}Xrrrrrrr@{}}
	\toprule
	Dataset                    & \# Sub.         & \# Chan. & \# Trials/class
	                           & \makecell{Epoch                                                                                          \\ len. (s)} & \makecell{S. freq.\\ (Hz)}
	                           & \# Sess.        & Ref.                                                                                   \\
	\midrule
	\textbf{\Ac{erp} datasets} &                 &          &                 &     &      &                   &                          \\
	BNCI2014-008               & 8               & 8        & 3500/700        & 1.0 & 256  & 1                 & \cite{Riccio2013}        \\
	BNCI2014-009               & 10              & 16       & 1440/288        & 0.8 & 256  & 3                 & \cite{Arico2014}         \\
	BNCI2015-003               & 10              & 8        & 1500/300        & 0.8 & 256  & 1                 & \cite{Guger2009}         \\
	BrainInvaders2012          & 25              & 16       & 640/128         & 1.0 & 128  & 2                 & \cite{VanVeen2019}       \\
	BrainInvaders2013a         & 24              & 16       & 3200/640        & 1.0 & 512  & 8 (sub. 1-7) or 1 & \cite{Vaineau2019}       \\
	BrainInvaders2014a         & 64              & 16       & 990/198         & 1.0 & 512  & up to 3           & \cite{Korczowski2019}    \\
	BrainInvaders2014b         & 38              & 32       & 200/40          & 1.0 & 512  & 3                 & \cite{Korczowski2019a}   \\
	BrainInvaders2015a         & 43              & 32       & 4131/825        & 1.0 & 512  & 3                 & \cite{Korczowski2019b}   \\
	BrainInvaders2015b         & 44              & 32       & 2160/480        & 1.0 & 512  & 1                 & \cite{Korczowski2019c}   \\
	Cattan2019-VR              & 21              & 16       & 600/120         & 1.0 & 512  & 2                 & \cite{Cattan2019}        \\
	EPFLP300                   & 8               & 32       & 2753/551        & 1.0 & 2048 & 4                 & \cite{Hoffmann2008}      \\
	Huebner2017                & 13              & 31       & 364/112         & 0.9 & 1000 & 3                 & \cite{Huebner2017}       \\
	Huebner2018                & 12              & 31       & 364/112         & 0.9 & 1000 & 3                 & \cite{Huebner2018}       \\
	Lee2019-ERP                & 54              & 62       & 6900/1380       & 1.0 & 1000 & 2                 & \cite{Lee2019}           \\
	Sosulski2019               & 13              & 31       & 7500/1500       & 1.2 & 1000 & 1                 & \cite{Sosulski2019}      \\
	\midrule
	\textbf{\Ac{mi} datasets}  &                 &          &                 &     &      &                   &                          \\
	AlexandreMotorImagery      & 8               & 16       & 20.0            & 3.0 & 512  & 1                 & \cite{Barachant2012}     \\
	BNCI2014-001               & 9               & 22       & 144.0           & 4.0 & 250  & 2                 & \cite{Tangermann2012}    \\
	Schirrmeister2017          & 14              & 128      & 120.0           & 4.0 & 500  & 1                 & \cite{Schirrmeister2017} \\
	Weibo2014                  & 10              & 60       & 80.0            & 4.0 & 200  & 1                 & \cite{Yi2014}            \\
	Zhou2016                   & 4               & 14       & 160.0           & 5.0 & 250  & 3                 & \cite{Zhou2016}          \\
	\bottomrule
\end{tabularx}

%% file: stats_erp.tex
\begin{tabular}{@{}lrrrrrr@{}}
	\toprule
	Decoder 1          & \multicolumn{4}{c}{BTTDA} & \multicolumn{2}{c}{PARAFACDA}                                                           \\
	Decoder 2          & \multicolumn{2}{c}{HODA}  & \multicolumn{2}{c}{PARAFACDA} & \multicolumn{2}{c}{HODA}                                \\
	                   & $p$                       & SMD                           & $p$                      & SMD  & $p$            & SMD  \\
	\midrule
	BNCI2014-008       & \num{3.91e-03}            & 1.57                          & \num{1.95e-02}           & 0.90 & \num{7.81e-03} & 1.31 \\
	BNCI2014-009       & \num{5.86e-03}            & 0.94                          & \num{1.76e-02}           & 0.80 & \num{5.47e-02} & 0.58 \\
	BNCI2015-003       & \num{2.93e-03}            & 1.31                          & \num{7.81e-03}           & 0.96 & \num{6.84e-03} & 1.11 \\
	BrainInvaders2012  & \num{6.14e-06}            & 1.74                          & \num{1.98e-02}           & 0.49 & \num{7.85e-06} & 1.45 \\
	BrainInvaders2013a & \num{3.28e-06}            & 1.05                          & \num{8.00e-02}           & 0.22 & \num{1.51e-05} & 0.98 \\
	BrainInvaders2014a & \num{2.56e-11}            & 1.18                          & \num{2.81e-03}           & 0.40 & \num{5.71e-11} & 1.14 \\
	BrainInvaders2014b & \num{3.02e-04}            & 0.62                          & \num{1.34e-01}           & 0.15 & \num{3.02e-04} & 0.59 \\
	BrainInvaders2015a & \num{1.59e-12}            & 1.17                          & \num{3.22e-07}           & 0.84 & \num{8.62e-10} & 1.00 \\
	BrainInvaders2015b & \num{2.11e-11}            & 1.33                          & \num{4.37e-02}           & 0.27 & \num{6.47e-10} & 1.19 \\
	Cattan2019-VR      & \num{6.68e-06}            & 1.28                          & \num{3.97e-01}           & 0.21 & \num{2.38e-06} & 1.38 \\
	EPFLP300           & \num{3.91e-03}            & 1.72                          & \num{3.52e-02}           & 0.81 & \num{3.91e-03} & 1.36 \\
	Huebner2017        & \num{3.66e-04}            & 0.62                          & \num{1.37e-01}           & 0.32 & \num{4.88e-04} & 0.61 \\
	Huebner2018        & \num{1.22e-03}            & 1.15                          & \num{3.91e-03}           & 0.88 & \num{3.17e-03} & 1.10 \\
	Lee2019-ERP        & \num{8.13e-11}            & 1.06                          & \num{3.30e-03}           & 0.38 & \num{1.08e-10} & 1.02 \\
	\bottomrule
\end{tabular}

%% file: stats_mi.tex
\begin{tabular}{@{}lrrrrrr@{}}
	\toprule
	Decoder 1          & \multicolumn{4}{c}{BTTDA} & \multicolumn{2}{c}{PARAFACDA}                                                           \\
	Decoder 2          & \multicolumn{2}{c}{HODA}  & \multicolumn{2}{c}{PARAFACDA} & \multicolumn{2}{c}{HODA}                                \\
	                   & $p$                       & SMD                           & $p$                      & SMD  & $p$            & SMD  \\
	\midrule
	BNCI2014-008       & \num{3.91e-03}            & 1.57                          & \num{1.95e-02}           & 0.90 & \num{7.81e-03} & 1.31 \\
	BNCI2014-009       & \num{5.86e-03}            & 0.94                          & \num{1.76e-02}           & 0.80 & \num{5.47e-02} & 0.58 \\
	BNCI2015-003       & \num{2.93e-03}            & 1.31                          & \num{7.81e-03}           & 0.96 & \num{6.84e-03} & 1.11 \\
	BrainInvaders2012  & \num{6.14e-06}            & 1.74                          & \num{1.98e-02}           & 0.49 & \num{7.85e-06} & 1.45 \\
	BrainInvaders2013a & \num{3.28e-06}            & 1.05                          & \num{8.00e-02}           & 0.22 & \num{1.51e-05} & 0.98 \\
	BrainInvaders2014a & \num{2.56e-11}            & 1.18                          & \num{2.81e-03}           & 0.40 & \num{5.71e-11} & 1.14 \\
	BrainInvaders2014b & \num{3.02e-04}            & 0.62                          & \num{1.34e-01}           & 0.15 & \num{3.02e-04} & 0.59 \\
	BrainInvaders2015a & \num{1.59e-12}            & 1.17                          & \num{3.22e-07}           & 0.84 & \num{8.62e-10} & 1.00 \\
	BrainInvaders2015b & \num{2.11e-11}            & 1.33                          & \num{4.37e-02}           & 0.27 & \num{6.47e-10} & 1.19 \\
	Cattan2019-VR      & \num{6.68e-06}            & 1.28                          & \num{3.97e-01}           & 0.21 & \num{2.38e-06} & 1.38 \\
	EPFLP300           & \num{3.91e-03}            & 1.72                          & \num{3.52e-02}           & 0.81 & \num{3.91e-03} & 1.36 \\
	Huebner2017        & \num{3.66e-04}            & 0.62                          & \num{1.37e-01}           & 0.32 & \num{4.88e-04} & 0.61 \\
	Huebner2018        & \num{1.22e-03}            & 1.15                          & \num{3.91e-03}           & 0.88 & \num{3.17e-03} & 1.10 \\
	Lee2019-ERP        & \num{8.13e-11}            & 1.06                          & \num{3.30e-03}           & 0.38 & \num{1.08e-10} & 1.02 \\
	\bottomrule
\end{tabular}